\documentclass[%
 reprint,
 amsmath,amssymb,
 aps,
pra,
]{revtex4-1}
\usepackage{float}
\usepackage{graphicx}% Include figure files
\usepackage{dcolumn}% Align table columns on decimal point
\usepackage{bm}% bold math
\usepackage{xcolor}
\usepackage[normalem]{ulem}
\newcommand{\ket}[1]{\left | #1 \right \rangle}

\definecolor{commentColorPAA}{rgb}{0.8,0.0,0.2}
\definecolor{commentColorMP}{rgb}{0.0,0.2,0.6}
\definecolor{commentColorAW}{rgb}{0.6,0.2,0.0}
\definecolor{commentColorASN}{rgb}{0.2,0.6,0.0}

\newcommand{\omegage}{\omega_\text{ge}}
\newcommand{\omegaef}{\omega_\text{ef}}
\newcommand{\omegam}{\omega_\text{m}}
\newcommand{\omegamone}{\omega_\text{m}^{(1)}}
\newcommand{\omegami}{\omega_\text{m}^{(i)}}

\newcommand{\opsigmaz}{\hat{\sigma}_z}
\newcommand{\opb}{\hat{b}}
\newcommand{\opbd}{\hat{b}^\dagger}
\newcommand{\op}[2]{{\hat{#1}_{#2}}}

\newcommand{\omegar}{\omega_\text{r}}
\newcommand{\kappar}{\kappa_\text{r}}
\newcommand{\Deltage}{\Delta_\text{ge}}
\newcommand{\Deltam}{\Delta_\text{m}}
\newcommand{\omegad}{\omega_\text{d}}

\newcommand{\opa}{\hat{a}}
\newcommand{\opad}{\hat{a}^\dagger}
\newcommand{\opc}{\hat{c}}
\newcommand{\opcd}{\hat{c}^\dagger}
\newcommand{\oprho}{\hat{\rho}}

\begin{document}

\preprint{APS/123-QED}

\title{Resolving the energy levels of a nanomechanical oscillator}% Force line breaks with \\

\author{Patricio Arrangoiz-Arriola}
\thanks{These authors contributed equally to this work}
\author{E. Alex Wollack}%
\thanks{These authors contributed equally to this work}
\author{Zhaoyou Wang}%
\author{Marek Pechal}%
\author{Wentao Jiang}%
\author{Timothy P. McKenna}%
\author{Jeremy D. Witmer}

\author{Amir H. Safavi-Naeini}%
 \email{safavi@stanford.edu}
\affiliation{%
 Department of Applied Physics and Ginzton Laboratory, Stanford University\\
 348 Via Pueblo Mall, Stanford, California 94305, USA
}%

\date{\today}

\pacs{Valid PACS appear here}
\maketitle

{\bf The quantum nature of an oscillating mechanical object is anything but apparent. The coherent states that describe the classical motion of a mechanical oscillator do not have well-defined energy, but are rather quantum superpositions of equally-spaced energy eigenstates. Revealing this quantized structure is only possible with an apparatus that measures the mechanical energy with a precision greater than the energy of a single phonon, $\hbar\omegam$.  One way to achieve this sensitivity is by engineering a strong but nonresonant interaction between the oscillator and an atom. In a system with sufficient quantum coherence, this interaction allows one to distinguish different phonon number states by resolvable differences in the atom's transition frequency. For photons, such dispersive measurements have been studied in cavity \cite{Brune1994, Bertet2002} and circuit quantum electrodynamics \cite{Schuster2007} where experiments using real and artificial atoms have resolved the photon number states of cavities. Here, we report an experiment where an artificial atom senses the motional energy of a driven nanomechanical oscillator with sufficient sensitivity to resolve the quantization of its energy. To realize this, we build a hybrid platform that integrates nanomechanical piezoelectric resonators with a microwave superconducting qubit on the same chip. We excite phonons with resonant pulses of varying amplitude and probe the resulting excitation spectrum of the qubit to observe phonon-number-dependent frequency shifts $\approx 5$ times larger than the qubit linewidth. Our result demonstrates a fully integrated platform for quantum acoustics that combines large couplings, considerable coherence times, and excellent control over the mechanical mode structure. With modest experimental improvements, we expect our approach will make quantum nondemolition measurements of phonons~\cite{Braginsky1996} an experimental reality, leading the way to new quantum sensors and information processing approaches~\cite{Ofek2016} that use chip-scale nanomechanical devices.}

In the last decade, mechanical devices have been brought squarely into the domain of quantum science through a series of remarkable experiments exploring the physics of measurement, transduction, and sensing \cite{Aspelmeyer2014, OConnell2010, Gustafsson2014a, Cohen2015, Riedinger2016a, Chu2018, Satzinger2018, Viennot2018}. Two paradigms for obtaining quantum control over these systems are {those} of cavity optomechanics (COM), where the position $\op{x}{}$ parametrically couples to a higher-frequency electromagnetic cavity~\cite{Aspelmeyer2014}, and quantum acoustics (QA), where an artificial atom or qubit exchanges quanta with a mechanical oscillator. The latter is the acoustic analog of cavity or circuit QED (cQED), the archetypal playground of quantum optics that has enabled a vast range of experiments {probing} quantum physics \cite{Mabuchi2002} and led to the emergence of the superconducting approach to quantum information processing \cite{Devoret2013}. In QA systems, the exchange of quanta between a qubit and a mechanical oscillator is described by the Hamiltonian $\hat{H}_{\text{int}} = g(\hat{\sigma}_- + \hat{\sigma}_+)(\opb + \opbd)$, where $g$ is the coupling rate, and $\hat{\sigma}_+\, (\hat{\sigma}_-)$ and $\opbd\, (\opb)$ are the raising and lowering operators of the qubit and mechanical modes, respectively. Strong coupling of the system is realized for $g$ greater than the decoherence rates of the qubit, $\gamma$, and mechanical mode, $\kappa$; in this limit, a single excitation can be resonantly swapped multiple times before being lost to the environment. Experiments operating in this regime have demonstrated quantum control of mechanical systems at the single phonon level \cite{OConnell2010, Chu2017,  Satzinger2018}, as well as preparation of higher Fock states using more elaborate protocols \cite{Chu2018}.

\begin{figure}[t]
    \centering
    \includegraphics[width=89mm]{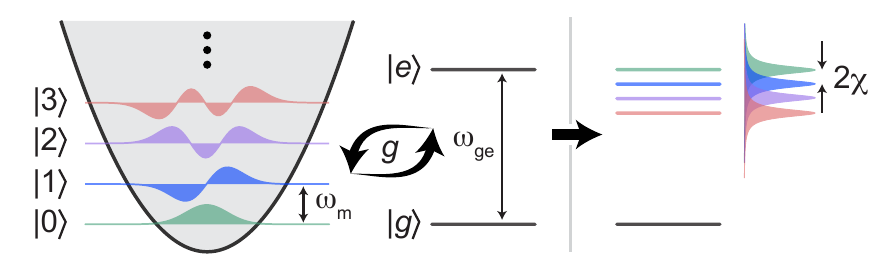}
    \caption{{\bf Phonon number splitting outline.}  The state of a mechanical oscillator is described in quantum mechanics by a linear superposition of equally-spaced {energy eigenstates $\ket{n}$, each representing a state of $n$ phonons in the system. This quantized structure is {normally} not {resolvable} since {the} transitions {between the energy levels} all occur at the same frequency {$\omegam$}. By coupling the resonator to a qubit {of transition frequency $\omegage$} with a rate $g$, we cause a splitting in the qubit spectrum  parameterized by {a dispersive coupling rate} $\chi$. {This}  allows us to distinguish between the different phonon number states {that are present in the oscillator}.}}
    \label{fig_schematic}
\end{figure}

In both COM and QA, approaches to probing the phonon number states of a mechanical resonator invariably involve swapping phonons into a resonator or qubit acting as a meter. Alternatively, we can build a measurement apparatus that directly senses the mechanical energy stored in a resonator without the need to exchange excitations~\cite{Viennot2018}. This quantum nondemolition (QND) approach to measuring motion has numerous advantages and remains an outstanding challenge in the study of mechanical systems in the quantum regime. Cavity optomechanical approaches to QND attempt to couple the detector to the $\op{x}{}^2$ as opposed to the $\op{x}{}$ observable of the mechanical system~\cite{Thompson2008}, but have required optomechanical coupling rates beyond current experimental capabilities~\cite{Miao2009,Ludwig2012} to achieve phonon number resolution. In QA, detuning the qubit transition frequency $\omegage = \omegam + \Delta$ from the mechanical frequency $\omegam$ by $|\Delta| \gg g$ prevents the direct swap of real excitations between the two systems. Instead, it leads to an off-resonant interaction between the qubit and mechanics that results in an energy-dependent shift of the qubit frequency induced by virtual transitions. The effective Hamiltonian
\begin{equation} \label{eq_dispersive_hamiltonian}
    \hat{H}_\text{eff} = \omegam \opbd \opb + \tfrac{1}{2}\left( \omegage + 2 \chi \opbd\opb \right) \opsigmaz,
\end{equation}
accurately describes the system in this off-resonant regime \cite{Schuster2007}. Under $\hat{H}_\text{eff}$, the only interaction between the two systems manifests as a qubit frequency shift $2\chi\hat{n}$ dependent on the phonon number  $\hat{n}=\opbd\opb$. For superconducting charge qubits operating in the transmon regime, the effects of the higher excited level ($\ket{f}$) must also be taken into account when calculating $\chi$. The resulting expression for the dispersive coupling rate $\chi$ is distinct from the two-level atom case, and given by \cite{Koch2007}
\begin{equation}
    \chi = -\frac{g^2}{\Delta}\frac{\alpha}{\Delta - \alpha},
\end{equation}
where $\alpha = \omegage - \omegaef$ is the transmon anharmonicity. Since $\hat{H}_\text{eff}$ commutes with both the phonon number operator $\hat{n}$ and $\opsigmaz$, the two systems cannot exchange energy, and so measurements of the qubit {excited state} population $(1 +  \opsigmaz)/2$ do not perturb the phonon number. Furthermore, in the limit $\chi \gg \text{max}\{\gamma , \kappa\}$, the frequency shift $2\chi$ induced by the presence of a single phonon in the oscillator becomes resolvable in the qubit excitation spectrum. We call this the phonon number splitting regime in analogy to the dispersive regime of cQED \cite{Brune1990, Schuster2007}, where photons in an electromagnetic cavity lead to an energy-dependent atomic transition frequency. In cQED, the dispersive regime has been instrumental in implementing new approaches to quantum measurement and error correction~\cite{Ofek2016}.  

\begin{figure}[t!]
    \centering
    \includegraphics[width=89mm]{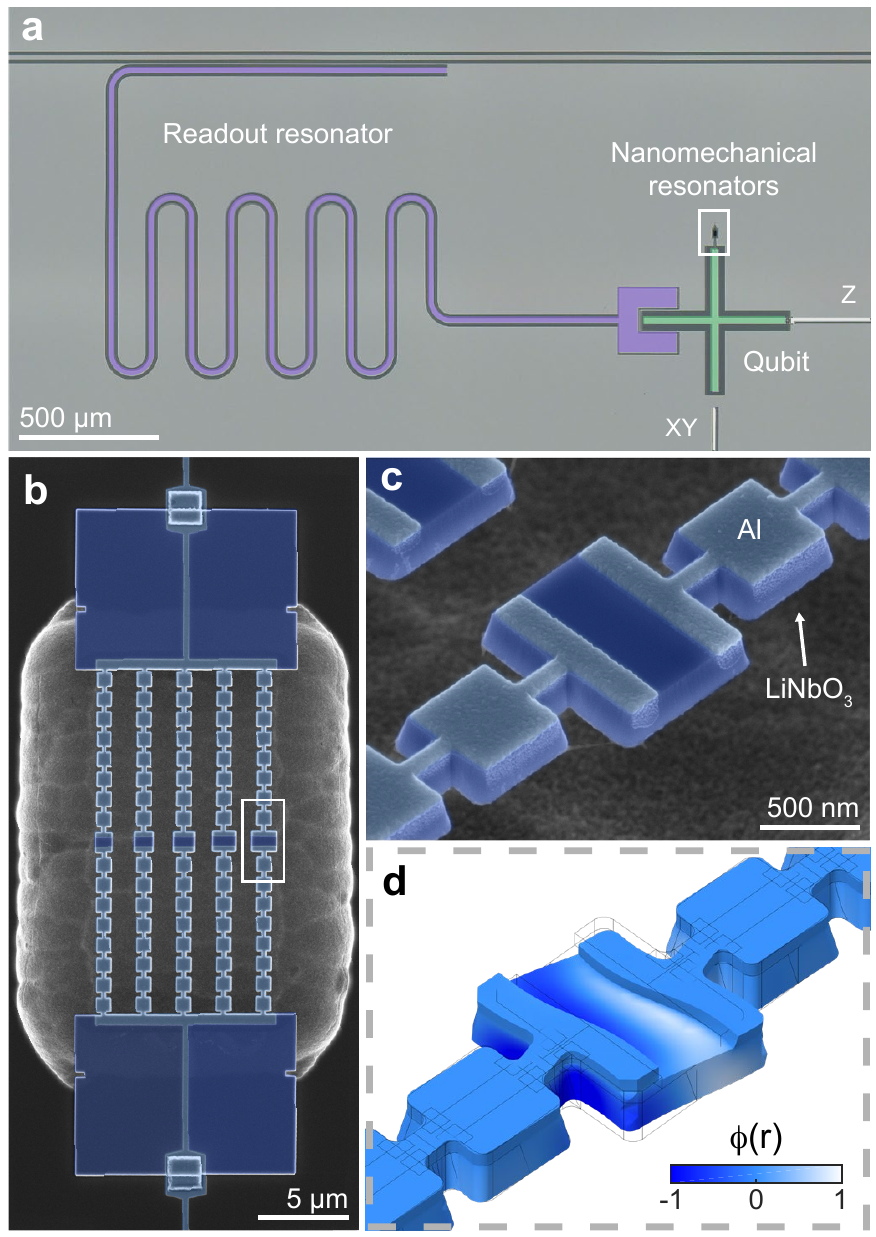}
    \caption{{\bf Device fabrication.} {\bf a,} Optical micrograph of the device showing the readout resonator (purple), transmon qubit (green), and nanomechanical resonators (white box). The qubit flux control ($Z$) and excitation ($XY$) lines are shown in white. {\bf b,} False-colored scanning-electron micrograph (SEM) of the suspended resonators. Each resonator consists of a defect site embedded in a phononic crystal that supports a complete phononic bandgap in the frequency range $\sim 2-2.4\,\text{GHz}$. The structures are fabricated from a $250 \,\text{nm}$ film of LN (dark blue) that is suspended above a silicon substrate, and are coupled to the qubit via thin aluminum electrodes (light blue) that address the defect modes. We form a connection between the electrodes and the qubit using superconducting bandages, which are visible as small squares at the edges of the LN supporting slabs. {\bf c,} SEM image of a phononic crystal defect. {\bf d,} Finite element simulation of a mechanical defect mode, showing the localized deformation of the structure and the electrostatic potential $\phi(\mathbf{r})$ (color) generated through the piezoelectricity of LN.}
    \label{fig_fab}
\end{figure}

Several technical hurdles have prevented phonon number splitting from being observed in quantum acoustics. The sub-micron wavelength of gigahertz-frequency acoustic phonons -- far smaller than the scale of the electrodes comprising the qubit circuits -- leads to an enormous phonon mode density accessible at the qubit transition frequency. Uncontrolled coupling to phonons is a known source of dissipation in cQED~\cite{Ioffe2004}, and represents a major challenge in combining qubits with the strong piezoelectrics needed for phonon sensing and control. Systems designed to have large coupling rates tend to couple strongly to parasitic modes, reducing the overall coherence of the qubits as well as the availability of viable operating frequencies. Approaches to mitigating these losses have included tunable couplers that isolate the qubit from the piezoelectric material by rapidly turning off the coupling after interaction~\cite{Satzinger2018}, and bulk wave resonators where the participation ratio of the qubit electric field with the piezoelectric is reduced~\cite{Chu2017,Chu2018}.  Our approach avoids sacrificing coupling, while maintaining qubit coherence. We reduce the density of accessible mechanical modes that the qubit can radiate into by confining the phonons to a very small piezoelectric region where only a few mechanical modes are present at the frequencies of interest. {Ordinarily}, the leakage of phonons out of this region through its supporting anchors would lead to rapid decoherence of both the mechanics and the qubit. To realize leakage-free anchors, we create a periodic patterning of the elastic material that opens a phononic bandgap. Equivalently, we can view the mechanical resonator as a defect in a phononic crystal bandgap material that is etched into a piezoelectric film.

To fabricate the chip-scale system, we integrate microwave Josephson junction qubits (aluminum on high-resistivity silicon) with piezoelectric nanomechanical devices patterned from thin-film lithium niobate (LN) (see Supplementary Information for fabrication details). As seen in Fig.~2, our system utilizes a transmon qubit of the type presented in Ref.~\cite{Barends2013}, controlled via on-chip microwave lines and read out dispersively through a microwave readout resonator. In turn, the transmon is coupled to an array of one-dimensional phononic crystal defect resonators through the piezoelectricity of LN. Each mechanical structure consists of a narrow, suspended beam of patterned LN (Fig.~2b) with a periodicity $a = 1\,\mu\text{m}$ that opens a complete bandgap in the $\sim 2 - 2.4\,\text{GHz}$ region. The defect site at the center of the phononic crystal supports highly confined mechanical modes with frequencies that lie within the bandgap (Fig.~2c,~d). In order to address these modes, we place aluminum electrodes directly on top of the phononic crystal anchors. With one terminal grounded and another terminal contacted to the transmon, the voltage fluctuations of the qubit create an electric field in the defect site which is linearly coupled to its mechanical deformation by the piezoelectric effect. The structure is designed such that at least one of the localized modes generates a polarization that is aligned with the electric field produced by the electrodes (see SI for design details). 

\begin{figure}[t]
    \centering
    \includegraphics[width=89mm]{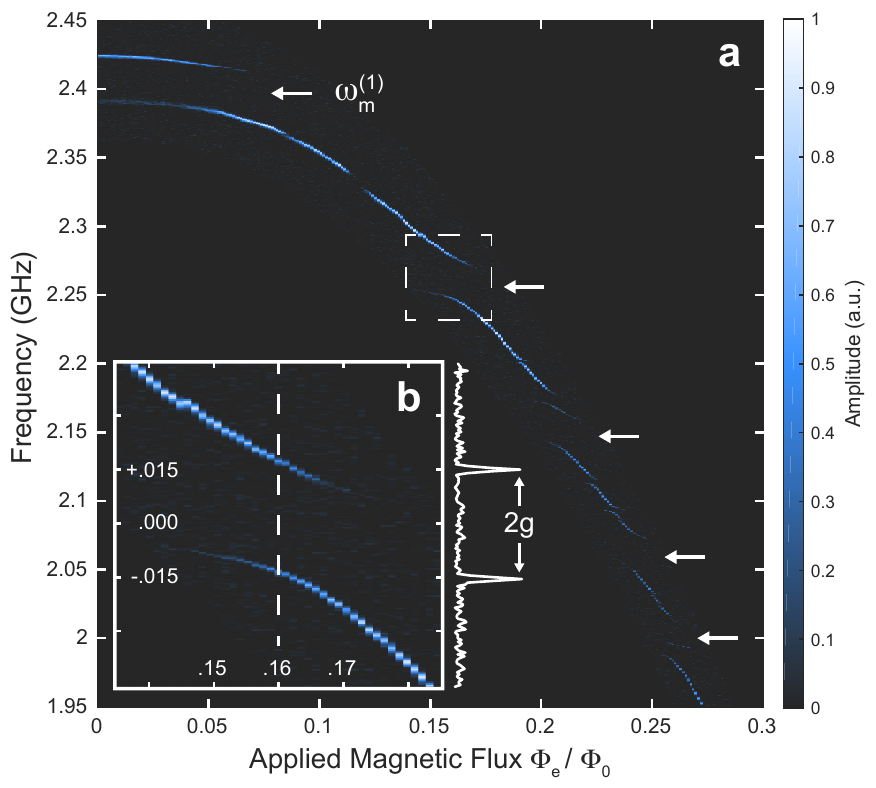}
    \caption{{\bf Qubit spectroscopy and mechanical mode structure.} {\bf a,} Qubit spectrum as a function of applied magnetic flux. Arrows indicate the strongly coupled mechanical modes associated with each of the five phononic crystal defect resonators. The qubit frequency at the flux sweet spot ($\Phi_e=0$) is $\omegage^{(\text{max})}/2\pi = 2.417\,\text{GHz}$, in close proximity to the highest-lying mechanical mode at $\omegamone/2\pi = 2.405\,\text{GHz}$ which was used for the phonon number splitting experiment. A small number of weakly coupled features are present in the spectrum, corresponding to additional localized defect modes. {\bf b,} Close-up of the anticrossing with the mechanical mode at $2.257\,\text{GHz}$ (dashed box in {\bf a}). The vertical slice at zero detuning is shown in white to the right, and is used to calculate a coupling rate $g / 2 \pi = 15.2\,\text{MHz}$.}
    \label{fig_tuning_curve}
\end{figure}

We first probe the mechanical resonances by measuring the qubit excitation spectrum as we tune its transition frequency $\omegage$ across the phononic bandgap region. Here, frequency control is provided by a magnetic flux applied via an on-chip flux line, and the qubit is excited using a dedicated charge line. The results of this measurement are shown in Fig.~\ref{fig_tuning_curve}a, where we observe a series of anticrossings corresponding to various defect modes. From this data we obtain the frequencies $\{\omegami\}$ and coupling rates $\{g_i \}$ of the five most strongly coupled modes, each corresponding to an individual resonator in the array. We measure coupling rates in the range $g/2\pi = 13 - 16\,\text{MHz}$, in fairly good agreement with finite-element simulations (see Supplementary Information). We also observe a set of anticrossings corresponding to a small number of additional, weakly coupled defect modes. For the phonon number splitting measurements presented later, we use the highest-lying mechanical mode at $\omegamone/2\pi = 2.405\,\text{GHz}$, for which we perform a ringdown measurement to find its decay rate $\kappa/2\pi = 370 \, \text{kHz}$. Next, in order to characterize the coherence of the qubit we tune it to $\omegage/2\pi = 2.301\,\text{GHz}$, sufficiently far from all mechanical modes, and measure a qubit energy relaxation time $T_1 = 1.14 \,\mu\text{s}$ and a total qubit linewidth $\gamma/2\pi \approx 600 \, \text{kHz}$. Finally, we extract the qubit anharmonicity $\alpha / 2\pi = 138 \,\text{MHz}$ using a two-tone spectroscopy measurement of the $|g\rangle \rightarrow |e\rangle$ and $|e\rangle \rightarrow |f\rangle$ transitions. All together, these parameters place the system deep in the strong-coupling regime ($g \gg \kappa,\gamma$), and open up the possibility of observing phonon number splitting, with an expected dispersive shift $2\chi/2\pi \approx 3\, \text{MHz}$.

\begin{figure}[t!]
    \centering
    \includegraphics[width=89mm]{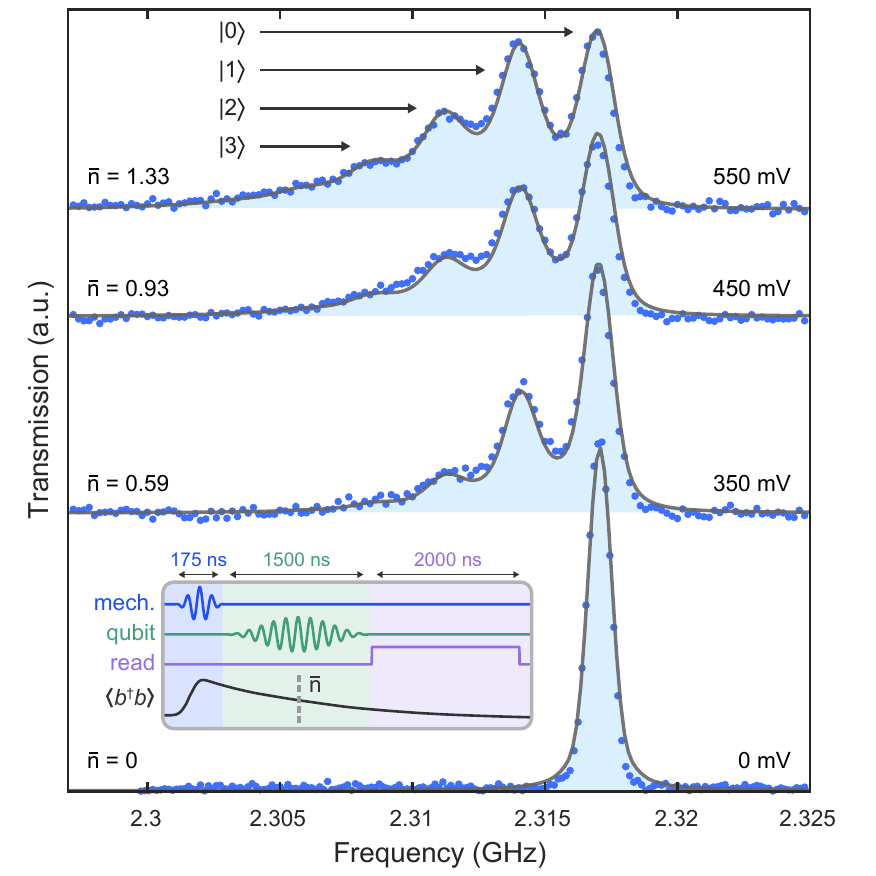}
    \caption{{\bf Phonon number splitting.} The qubit excitation spectrum is measured following a phonon excitation pulse of duration $\tau_\text{mech} = 175 \, \text{ns}$ and varying amplitude (see inset for pulse sequence). The initial phonon populations prepared by the pulse decay over the course of the measurement, but are nevertheless visible as individual peaks split by twice the dispersive coupling rate, $2\chi$. At the highest drive amplitudes we are able to resolve up to the $n=3$ phonon number state. We fit the data (blue points) using numerical master equation simulations of the full pulse sequence (solid black lines), with the mechanical drive strength as the only free fit parameter in the Hamiltonian. From these simulations we extract the mean phonon number $\bar{n} := \langle \hat{n}(\tau_\text{mech} + \tau/2) \rangle$ midway through the qubit spectroscopy pulse, which we indicate next to each spectrum.}
    \label{fig_number_splitting}
\end{figure}

In order to observe phonon number splitting, we perform a pump-probe measurement consisting of a short phonon excitation pulse followed by a longer qubit spectroscopy pulse (Fig.~\ref{fig_number_splitting}~inset). The phonon pulse is sent to the $XY$ line of the qubit. Since the qubit and mechanics are weakly hybridized {when far detuned}, the pulse drives the mechanical system into an approximate coherent state. The duration $\tau$ of the spectroscopy pulse was chosen to balance two competing effects. The pulse bandwidth $\sim 1/\tau$ needs to be sufficiently small in order to resolve the narrowest spectroscopic features, which are of width $\sim \gamma$ in our system, while $\tau$ cannot be much longer than the phonon lifetime $1/\kappa$ since the mechanical mode must remain excited during measurement. Of the two requirements $\tau \gg \gamma^{-1}=270\,\text{ns}$ and $\tau \lesssim \kappa^{-1}=430\,\text{ns}$, the first is necessary to observe number splitting, while the second determines the effective size of the observed mechanical state. We choose $\tau = 1.5\,\mu\text{s}$, which satisfies the first but not the second condition, in order to obtain better resolution for the phonon number peaks. As we perform the measurement, the mechanical mode experiences significant dissipation, which limits the mean number of phonons we can observe in this experiment to $\langle \hat{n} \rangle$ $\sim 1$.

We use the highest-lying mechanical mode at $\omegamone / 2\pi = 2.405\,\text{GHz}$ and detune the qubit by $\Delta \approx -6 g$ to $\omegage /2\pi = 2.317\,\text{GHz}$. By varying the amplitude of the preparation pulse, we prepare states of varying phonon occupations, resulting in the qubit spectra found in Figure \ref{fig_number_splitting}. In addition to the $|g\rangle \to |e\rangle$ qubit transition, we observe a series of peaks corresponding to different phonon number states $|n\rangle$ populated by the preparation pulse. The peaks are uniformly separated by $2\chi/2\pi \approx 3 \, \text{MHz}$, in close agreement with the dispersive shifts expected with our device parameters. The amplitude of the $\text{n}^\text{th}$ peak is an indirect measure of the population of state $|n\rangle$, as evidenced by the fact that the relative heights of the peaks associated with $n>0$ increase at higher excitation voltages. We also observe phonon number dependent linewidths for each peak, which can be understood as dephasing of the qubit due to the more rapid decay $n\kappa$ of higher-lying Fock states \cite{Gambetta2006}. This broadens higher phonon number peaks and obscures the quantization of the mechanical oscillator's energy. Therefore, at sufficiently {large} phonon occupations we enter a regime where the effect of the mechanical motion on the qubit spectrum is that of an ac-Stark shift induced by a coherent field \cite{Schuster2005, Viennot2018}. 

We numerically model our measurement using time domain master equation simulations that evolve the joint state $\hat{\rho}(t)$ of the mechanical mode and qubit with the full Hamiltonian over the course of the pulse sequence (see Supplementary Information). The state of the system  at the end of the excitation and spectroscopy pulses is given by $\hat{\rho}_\text{f} := \hat{\rho}(\tau_\text{mech} + \tau)$ and is used to calculate the qubit excited state populations $p_\text{e} = \text{Tr}\{\hat{\rho}_\text{f}|e\rangle \langle e|\}$. These are overlaid with the data in Fig.~\ref{fig_number_splitting}. The parameters used in the simulation are obtained from an independent set of calibrations as described in the Supplementary Information. The only free parameter is a correction factor (on the order of 1) for the mechanical drive strength. An offset and a scaling factor are used to overlay the simulated excitation spectrum on the measurements. To provide an approximate measure of the size of the mechanical states in the resonator, we indicate the mean phonon number $\bar{n} := \text{Tr}\{\hat{\rho}(\tau_\text{mech} + \tau/2)|n\rangle \langle n|\}$ midway through the spectroscopy pulse next to each spectrum in Fig.~\ref{fig_number_splitting}. 

We have demonstrated a quantum acoustic platform that combines phononic crystal defect modes with superconducting qubits.  
By using a phononic crystal bandgap, we reduce the mechanical and qubit dissipation rates while maintaining a large phonon-qubit coupling $g$. This enables us to dispersively resolve the phonon number states of a mechanical resonator -- a key step towards {realizing} QND measurements of a solid mechanical object and detecting quantum jumps of phonon {number}~\cite{Braginsky1996}. Looking forward, we expect phononic-crystal-based quantum acoustics to enable a new class of hybrid quantum technologies and provide a natural platform for integrating strongly piezoelectric materials with superconducting qubits. These types of mechanical resonators are also naturally suited for efficient optical readout due to the large mechanical mode confinement, and can provide a route for networking of microwave quantum machines~\cite{Safavi-Naeini2011,Bochmann2013}. Moreover, extremely long coherence times on the order of $1.5~\text{s}$ have now been demonstrated on phononic crystal devices implemented in silicon~\cite{MacCabe2019}, suggesting that the mechanical dissipation of our devices can be improved with further investigation. Ultra-coherent mechanical resonators integrated with qubits provide a route to realizing quantum acoustic processors where phononic registers act as quantum memories that {may} simplify scaling of superconducting quantum machines~\cite{Pechal2018}. Finally, by moving into the strong dispersive regime, our work enables further demonstrations, such as quantum nondemolition detection of single phonons \cite{Braginsky1996} and generation of Schr{\"o}dinger cat states of motion \cite{Vlastakis2013}.

\section*{Acknowledgments}

The authors would like to thank R. Patel, C. J. Sarabalis, R. Van Laer, and A. Y. Cleland for useful discussions. This work was supported by the David and Lucille Packard Fellowship, the Stanford University Terman Fellowship, and by the U.S. government through the Office of Naval Research under MURI No. N00014-151-2761 (QOMAND), and by the National Science Foundation under grant No. ECCS-1708734. P.A.A. and J.D.W. were partially supported by a Stanford Graduate Fellowship, and E.A.W. was supported by the Department of Defense through the National Defense \& Engineering Graduate Fellowship. M.P ackowledges support from the Swiss National Science Foundation. Part of this work was performed at the Stanford Nano Shared Facilities (SNSF), supported by the National Science Foundation under Grant No. ECCS-1542152, and the Stanford Nanofabrication Facility (SNF). 

\section*{Author contributions}

P.A.A. and E.A.W. designed and fabricated the device. P.A.A., E.A.W., W.J., T.P.M, and J.D.W. developed the fabrication process. Z.W., M.P., and A.-H.S.N. provided experimental and theoretical support. P.A.A. and E.A.W. performed the experiments and analyzed the data. P.A.A., E.A.W., and A.-H.S.N. wrote the manuscript, with Z.W. and M.P. assisting. P.A.A. and A.S.N conceived the experiment, and A.-H.S.N. supervised all efforts.

\section*{Author information}

The authors declare no competing financial interests. All correspondence should be addressed to A. H. Safavi-Naeini (safavi@stanford.edu).

%%%%%%%%%%%%%%%%%%%%%%%%%%%%%%%%%%%%%%%%%%%%%%%%
%%%%%%%%%%%%%%%%%%%%%%%%%%%%%%%%%%%%%%%%%%%%%%%%
% Supplementary Information
%%%%%%%%%%%%%%%%%%%%%%%%%%%%%%%%%%%%%%%%%%%%%%%%
%%%%%%%%%%%%%%%%%%%%%%%%%%%%%%%%%%%%%%%%%%%%%%%%

\pagebreak

\widetext
\begin{center}
\textbf{\large Supplemental information for: Resolving the energy levels of a nanomechanical oscillator}
\end{center}

\section{Device fabrication and properties}

\subsection{Fabrication details}

Our fabrication process begins with a $500\,\text{nm}$ film of lithium niobate (LN) on a $500\,\mu \text{m}$ high-resistivity ($\rho>\,3\,\text{k}\Omega\,\cdot\,\text{cm}$) silicon substrate. The LN film is first thinned to approximately 250\,nm by blanket argon milling. We then pattern a mask on negative resist (HSQ) with electron-beam (e-beam) lithography and transfer it to the LN with an angled argon milling step \cite{Wang2014}. After stripping the resist, we perform a thorough acid clean in order to remove re-deposited amorphous LN. This is critical, as any remaining residue significantly lowers the quality of the electrodes deposited in a later step. Next, we define the aluminum ground plane, feedlines, and transmon capacitor on the exposed silicon substrate with photolithography, electron beam evaporation, and liftoff. The Al/AlO$_\text{x}$/Al Josephson junctions are then formed using a standard Dolan bridge technique and double-angle evaporation \cite{Dolan1977, Kelly2015}. Following junction growth, we deposit 50\,nm aluminum electrodes directly on top of the phononic crystals in order to couple the defect modes to the qubit. This liftoff mask is patterned using e-beam lithography with $\sim10\,\text{nm}$ alignment precision to the existing LN structures. In the final metallization step, we evaporate aluminum bandages which form superconducting connections between the qubit capacitor, electrodes, junctions and ground plane \cite{Dunsworth2017}. The bandages are 500\,nm thick in order to smoothly connect the phononic crystal electrodes --- resting on the 250\,nm LN film --- with the qubit capacitor and ground plane below. After dicing the sample into individual chips, the LN structures are released with a masked $\text{XeF}_2$ dry etch that attacks the underlying silicon with high selectivity \cite{Vidal-alvarez2017}. Finally, the release mask is stripped in solvents and individual chips are packaged for low-temperature measurement.

\subsection{Device parameters}
Table \ref{table_device_params} gives device parameters for the qubit, the five strongly coupled mechanical modes, and the coplanar waveguide readout resonator. The maximum qubit frequency $\omegage^{(\text{max})}$ is extracted from a fit to the flux tuning curve, $\omegage(\Phi_e) = \omegage^{(\text{max})} \sqrt{|\cos(\pi \Phi_e / \Phi_0)|}$,  where $\Phi_e$ is the externally applied magnetic flux and $\Phi_0$ is the magnetic flux quantum. The transmon anharmonicity $\alpha  = \omegage - \omega_{\text{ef}}$ can also be extracted from the flux tuning curve, and we confirm this value with a separate two-tone measurement of the $|g\rangle\rightarrow |e\rangle$ and $|e\rangle\rightarrow |f\rangle$ transitions. Additionally, the qubit is characterized by its energy relaxation time, $T_1$, and its total linewidth, $\gamma$. These parameters were measured using  techniques  described in Section \ref{sec_rabi}  from which we obtain an estimate for the dephasing time $T_\phi$ through $\gamma = 1/2T_1 + 1/T_\phi$. The five strongly coupled mechanical modes are characterized by their resonant frequencies $\{\omegami\}$ and their coupling rates $\{g_i\}$ to the qubit, which are obtained by measuring the normal mode splittings in the flux tuning dataset (Fig.~3a of the main text). We note that the extraction of $g_3$ and $g_5$ is complicated by the presence of additional weakly coupled modes. The decay rate $\kappa$ of the mechanical mode $\omegamone$ (used for the number splitting experiment) was obtained via a ringdown measurement similar to a qubit $T_1$ measurement, as described in Section \ref{sec_rabi}. Finally, $\chi$ is the dispersive coupling rate between the transmon and the mechanical mode at the detuning $\Delta = \omegage - \omegam^{(1)}$ used in the experiment, $\Delta/2\pi = -88\,\text{MHz}$. We also list the frequency $\omega_\text{r}$ and linewidth $\kappar$ of the readout resonator.

\begin{table}[t]
    \centering
    \begin{tabular}{c c} \hline
        Parameter & Value \\ \hline
        
        $\omegage^{(\text{max})} /2\pi$ & $2.417\,\text{GHz}$ \\
        $\alpha / 2\pi$ & $138\,\text{MHz}$ \\
        $T_1$ & $1.0 - 1.4 \,\mu\text{s}$ \\
        $\gamma / 2\pi$ & $600 \,\text{kHz}$ \\
        $\{\omegami\} /2\pi$ & $2.405, 2.257, 2.153, 2.065, 2.002\,\text{GHz}$ \\
        $\{g_i\}/2\pi$ & $15.7, 15.2, \sim14, 14.2, \sim 13\,\text{MHz}$ \\
        $\kappa/2\pi$ & $370 \,\text{kHz}$ \\
        $\chi/2\pi$ & $-1.56\,\text{MHz}$\\
        $\omegar/2\pi$ & $3.026\,\text{GHz}$\\
        $\kappar/2\pi$ & $1.3\,\text{MHz}$\\ 
        \hline
    \end{tabular}
    \caption{Device parameters for the qubit, strongly coupled mechanical modes, and readout resonator.}
    \label{table_device_params}
\end{table}

\begin{figure*}[t!]
    \centering
    \includegraphics[width=178mm]{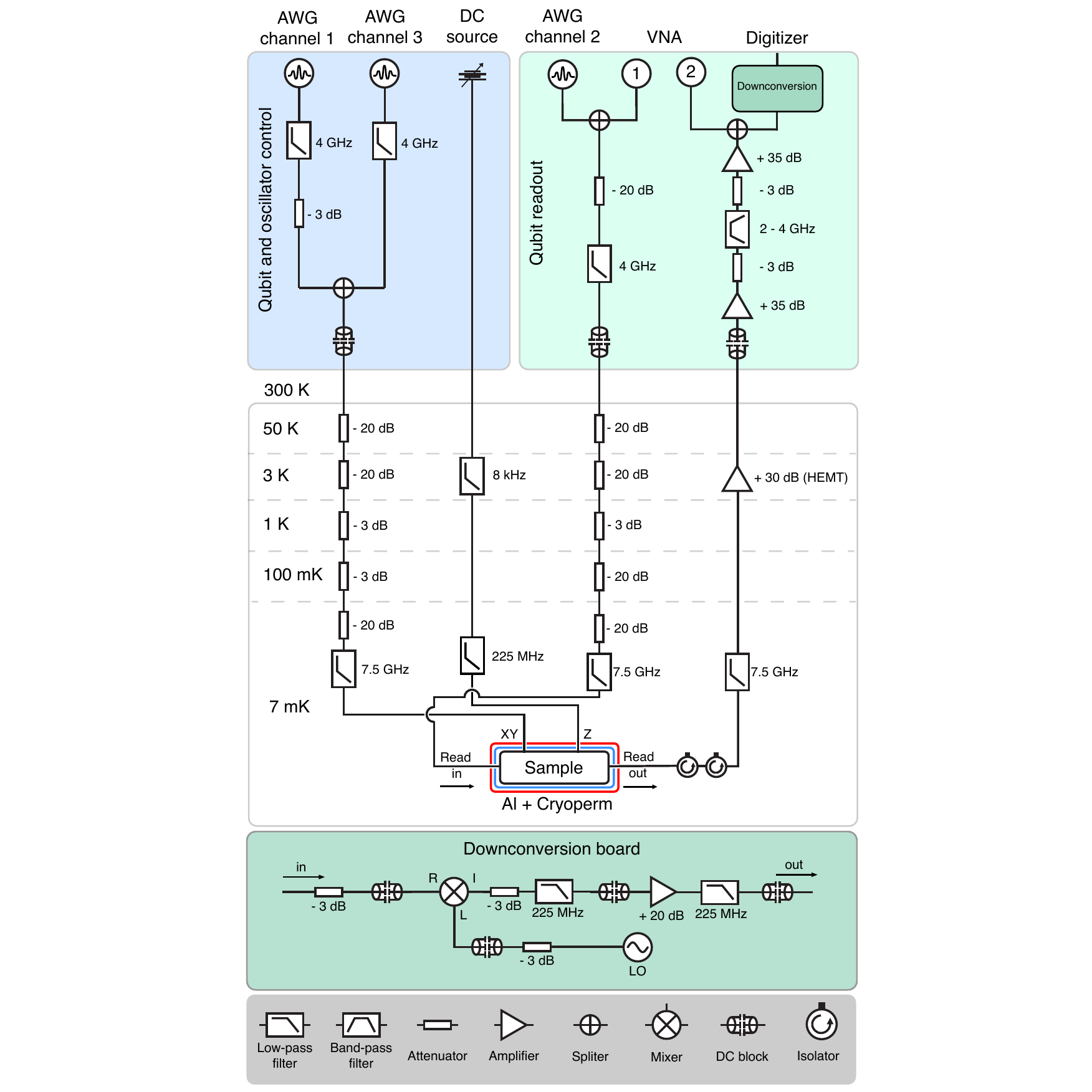}
    \caption{{\bf Experimental setup.} The sample is located at the mixing-chamber plate of a dilution refrigerator, packaged in a microwave PCB and copper enclosure, and surrounded by cryogenic magnetic shielding. All instruments are phase-locked by a 10\,MHz rubidium frequency standard (SRS SIM940).} 
    \label{fig_SI_setup}
\end{figure*}

\section{Experiment details}
\subsection{Qubit and mechanical oscillator control} \label{sec_qubit_control}
The experimental setup is shown in Fig.~\ref{fig_SI_setup}. In these experiments, we generate all qubit and phonon excitation pulses using a 5\,GS/s arbitrary waveform generator (AWG) (Tektronix series 5200). Because the qubit has a relatively low transition frequency ($\omegage/2\pi \approx 2.4\,\text{GHz}$), the pulses are produced directly using the instrument's built-in digital IQ mixer without further need for upconversion. The AWG output is then low-pass filtered at room temperature to remove Nyquist images, spurious intermodulation signals, and clock bleedthrough. We use a separate AWG channel to generate the phonon excitation pulses, which are then combined with the qubit pulses at room temperature. Once in the cryostat, the signals are attenuated and filtered at various temperature stages before being routed to the qubit through a dedicated charge line on the device (labeled $XY$ in Fig.~\ref{fig_SI_setup}). Flux biasing is performed using a programmable voltage source (SRS SIM928), which is low-pass filtered at the 3\,K stage (Aivon Therma-24G) and at the 7\,mK stage; the DC signal is then sent to an on-chip flux line (labeled $Z$ in Fig.~\ref{fig_SI_setup}). 

\subsection{Qubit readout}
The qubit state is read out dispersively via a  superconducting coplanar waveguide resonator \cite{Wallraff2004}. Square-envelope readout pulses are generated directly by the AWG with a carrier frequency of $\omegar /2\pi = 3.026\,\text{GHz}$, roughly matching the resonance frequency of the readout resonator. One end of the resonator is capacitively coupled to the qubit, while the other end is inductively coupled to a through-feedline with a coupling rate $\kappar / 2\pi = 1.3\,\text{MHz}$. After passing through two isolators (Quinstar QCY-030150S000), the signal is amplified at 3\,K by a high electron mobility transistor (HEMT) amplifier (Caltech CITCRYO1-12A), and at room temperature by two low-noise amplifiers (Miteq AFS4-02001800-24-10P-4 and AFS4-00100800-14-10P-4). Next, the signal is down-converted to an intermediate frequency (IF) of 125\,MHz using a separate local oscillator (Keysight E8257D), and a double-balanced mixer (Marki ML1-0220I). Finally, the IF signal is amplified, low-pass filtered, and digitized by an acquisition card (AlazarTech ATS9350) with 12-bit resolution and a 500\,MS/s sampling rate. The data is first stored on-board and then transferred to a GPU for real-time processing. Additionally, a vector network analyzer (Rhode \& Schwarz ZNB20) is used in the readout chain in order to calibrate the frequency of the readout pulses. 

\subsection{Characterization of qubit and mechanics}
\label{sec_rabi}
\begin{figure}[t!]
    \centering
    \includegraphics[width=89mm]{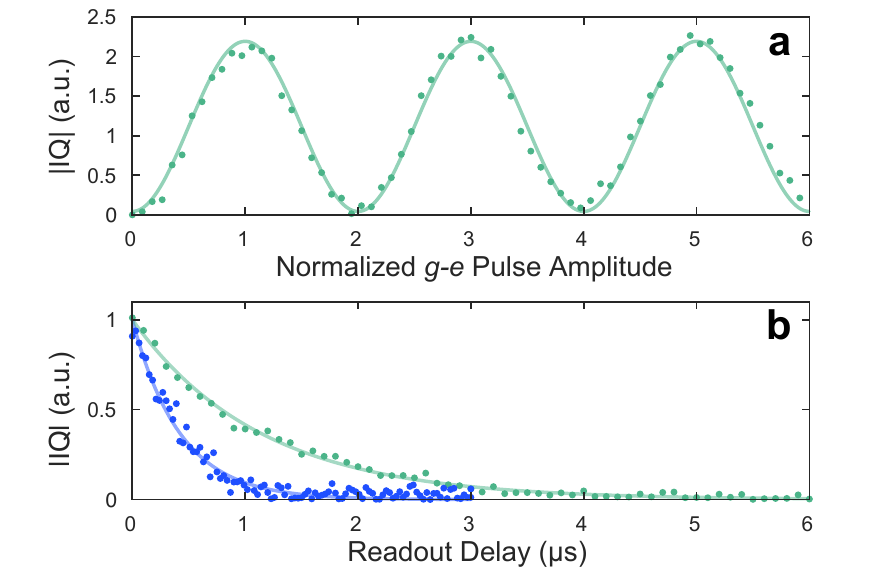}
    \caption{{\bf Characterization of qubit and mechanical modes.} {\bf a,} Rabi oscillation measurement used to calibrate the qubit excitation pulses. {\bf b,} Energy relaxation of qubit (green) and mechanics (blue), with lifetimes $T_1 = 1.14\,\mu\text{s}$ and $430\,\text{ns}$, respectively. This measurement of the qubit $T_1$ time was obtained at $\omegage/2\pi = 2.301\,\text{GHz}$. At this same frequency, we perform the ringdown measurement of the mechanical mode at $\omegamone = 2.405\,\text{GHz}$.}
    \label{fig_SI_ringdown}
\end{figure}

As described in Section \ref{sec_numerical_sims}, the system is driven through the transmon $XY$ line at a time-dependent Rabi rate $\Omega(t) = \Omega_0 f(t)$, where $f(t)$ is a normalized pulse envelope and $\Omega_0 = A_k V_\text{d}$ is directly proportional to the drive voltage $V_\text{d}$. The conversion factor $A_k$, with $k=1,3$ depends on which AWG channel is used to drive the qubit (see Fig.~\ref{fig_SI_setup}), and varies with frequency. We first calibrate the qubit excitation pulses through a Rabi oscillation measurement with the qubit at $\omegage/2\pi = 2.318 \, \text{GHz}$, the frequency at which we performed the number splitting measurements (see Fig.~\ref{fig_SI_ringdown}a). Here we use gaussian pulses $f(t) = \exp(-t^2 / 2\sigma_t^2)$ of width $\sigma_t = 50\, \text{ns}$ and varying amplitude $\Omega_0(V_\text{d})=A_1 V_\text{d}$. From this data, we extract $A_1 = 2\pi \times 93.9 \,\text{MHz} / \text{V}$. We infer $A_3 \approx \sqrt{2} A_1$ from the presence of the extra $3\,\text{dB}$ attenuator.

In order to measure the qubit energy relaxation time, $T_1$, we use the calibration to choose an appropriate $\pi$-pulse amplitude and approximately prepare the qubit in the excited state, $\ket{e}$. We then measure the excited state population (Fig.~\ref{fig_SI_ringdown}b) as we vary the delay between preparation and readout. The resulting data is fit to an exponential to extract $T_1$. We perform this measurement at a variety of qubit frequencies, all sufficiently separated from the strongly coupled mechanical modes, and measure relaxation times in the range $T_1 = 1.0 - 1.4\,\mu\text{s}$. In addition, we perform steady-state spectroscopy with the qubit at the frequency of the number splitting experiment in order to extract the total qubit linewidth, $\gamma/2\pi \approx 600 \, \text{kHz}$.

We perform a ringdown measurement to extract the decay rate $\kappa$ of the mechanical mode used for the phonon number splitting experiment. Here, the qubit is first detuned by an amount $\Delta \gg g$ in order to avoid hybridizing the modes (see Section \ref{sec_phonon_excitation}), and we then send a nearly resonant pulse at frequency $\omegad \approx \omegamone$ in order to excite the mechanical mode. The mean mechanical occupation $\langle \opbd \opb \rangle$ shifts the readout resonator through a small cross-Kerr interaction induced by the qubit. We therefore use this shift as an approximate measure of the occupation in the same way the excited state population of the qubit is measured. Sweeping the delay between excitation and readout produces the ringdown curve shown in Fig.~\ref{fig_SI_ringdown}b, with a decay rate $\kappa/2\pi = 370\,\text{kHz}$ corresponding to an energy relaxation time $\kappa^{-1} = 430\,\text{ns}$. 

All the parameter estimates obtained from these characterization measurements are later used in the numerical simulations of the pulse sequence, presented in Section~\ref{sec_numerical_sims}.

\subsection{Pulse sequence}
\label{sec_pulse_seq}
The phonon number splitting data was obtained using a pump-probe scheme in which a short phonon excitation pulse is sent at the mechanical frequency $\omegam$ and is immediately followed by a weak spectroscopy pulse. As described in Section \ref{sec_qubit_control}, these pulses are generated with separate AWG channels and are later combined before entering the cryostat. Both pulses have cosine-shaped envelopes of the form $V(t) = V_0[1 - \cos(2\pi t/ \tau)]/2$, which are synthesized at a baseband frequency of $\nu_\text{IF} = 125\, \text{MHz}$ and then digitally upconverted to their final carrier frequencies. For all the number splitting measurements, the length of the phonon excitation pulse is held fixed at $\tau_{\text{mech}} = 175 \, \text{ns}$, while its voltage is varied to prepare states with different mean phonon numbers $\langle \opbd \opb \rangle$. The length and voltage of the qubit spectroscopy pulse are respectively set to $\tau = 1.5 \, \mu\text{s}$ and $V_0 = 7.5 \, \text{mV}$, corresponding to a Rabi rate $\Omega_0 / 2\pi \approx 700 \, \text{kHz}$ (see Section~\ref{sec_rabi} for details on pulse calibration). We separately verified that these pulse settings do not result in power broadening of the qubit line. Immediately following the spectroscopy pulse, a $2 \, \mu\text{s}$ square-envelope pulse is sent to the readout port of the device. We measure the I and Q quadratures of the scattered pulse and subtract the reference values $I_0$ and $Q_0$ recorded with the system in the ground state. The resulting signal is then an indirect measure of the excited state population.

\begin{figure*}[t!]
    \centering
    \includegraphics[width=178mm]{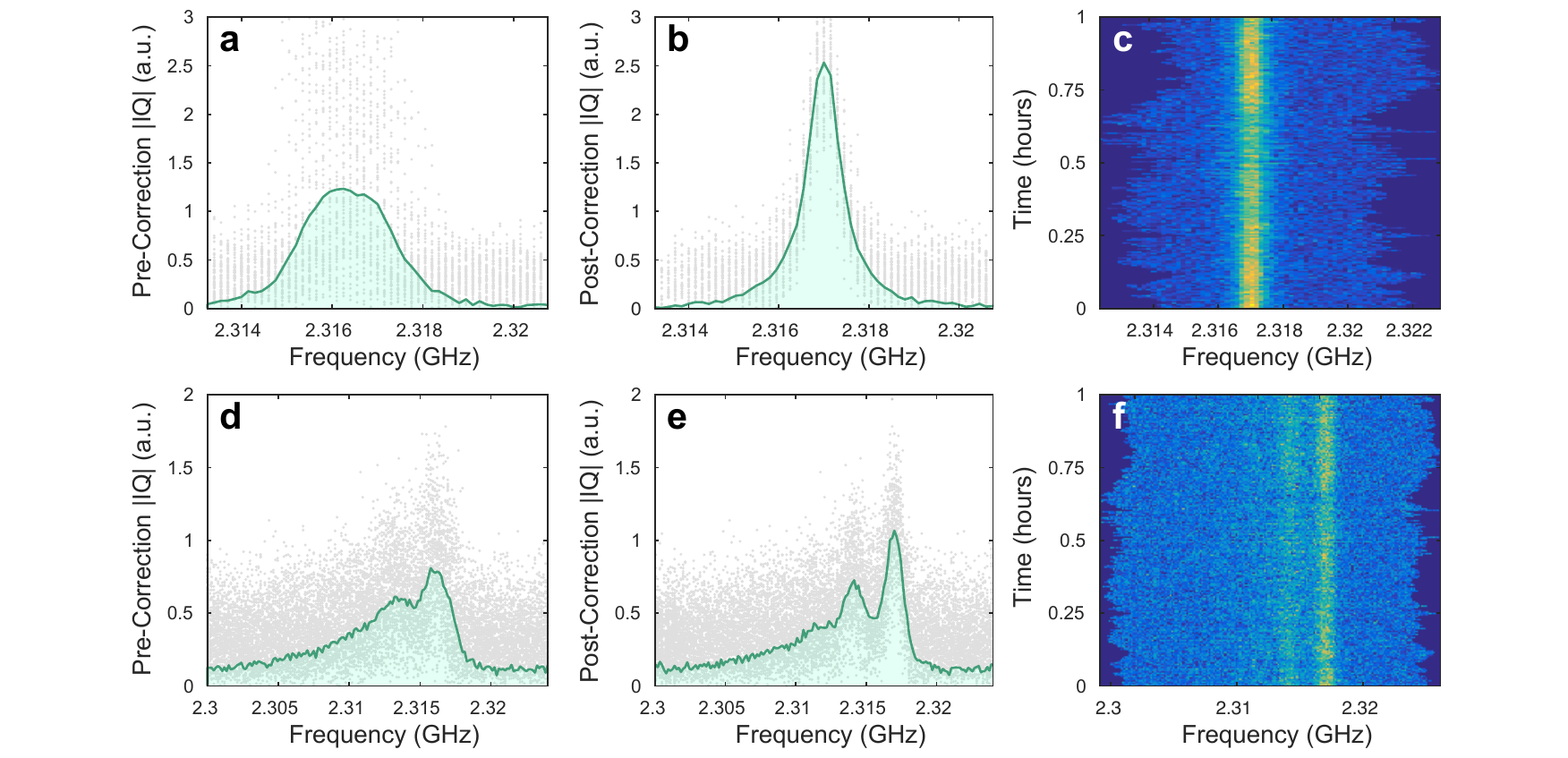}
    \caption{{\bf Qubit frequency tracking.} {\bf a,} Raw qubit tracking spectrum, 1 hour elapsed time. First, the bare qubit spectrum is averaged over a 20\,s interval, with a full spectral measurement performed every $\sim 1\,\text{ms}$; all of the raw 20\,s tracking spectra taken during the hour-long experiment are overlaid (gray circles). Then these spectra are averaged without post-processing to obtain the effective bare qubit spectrum (green curve), resulting in an effective linewidth $\sim 2.8\,\text{MHz}$. {\bf b,} Qubit tracking spectrum with post-processing. Here, we show the same qubit spectra as found in (a), but each 20\,s tracking spectrum is aligned to the average qubit frequency using post-processing peak detection. The effective qubit linewidth is now improved to $\sim 1.1\,\text{MHz}$. {\bf c,} Alignment of bare qubit spectra through time. Each horizontal slice represents a 20\,s tracking spectrum, showing that the qubit drifts by $\pm 1.5\,\text{MHz}$ during the experiment. {\bf d,} Raw phonon number splitting spectrum. We interleave number splitting measurements between the tracking spectra taken in (a), alternating between the two every $\sim 0.5\,\text{ms}$. When all of the raw spectra (grey points) are averaged without frequency correction (green curve), number splitting is visible but the peaks are poorly resolved. {\bf e,} Post-processed phonon number splitting spectrum. Using the frequency corrections calculated in (b), we adjust the frequencies of each slice of (d) and improve the resolution of the peaks. {\bf f,} Alignment of phonon number splitting spectra through time. The zero- and one-phonon peaks are easily visible with a splitting $2\chi \approx 3\,\text{MHz}$.}
    \label{fig_SI_qubit_tracking}
\end{figure*}

\subsection{Flux drift correction and qubit tracking}
Our experiment uses a frequency-tunable transmon qubit that is tuned away from its flux sweet spot, making it susceptible to flux noise and drift. We take several precautions to reduce drift in the qubit frequency induced by variations in the environmental magnetic field. These include low-temperature magnetic shielding, vibrational isolation, and low-pass filtering the DC flux bias. Nonetheless, some of the data presented took more than one hour of averaging to record. The slow drift in the qubit frequency, on the order of the linewidth of the qubit over one hour, smears out the peaks. We correct the slow qubit frequency drift by post-processing the data. 

The drift correction scheme is accomplished by alternating between measurements of the qubit frequency and the number splitting spectrum (Fig.~\ref{fig_SI_qubit_tracking}), and using the qubit frequency as a reference to align the number splitting data for averaging. This is done by creating a two-part AWG sequence: in the first part of the sequence, we measure the qubit spectroscopic line in a narrow window around the expected (noise-less) qubit frequency while the mechanical system is left unexcited. The next part of the sequence performs the number splitting measurement, i.e. phonon excitation followed by qubit spectroscopy. The resulting IQ averages are returned for data collection every $20\,\text{s}$. In post-processing, the frequency drift is extracted from the qubit tracking spectrum, then used to offset the number splitting data.  This scheme is able to reduce the apparent qubit linewidth from $2.8\,\text{MHz}$ to $1.1\,\text{MHz}$ during an hour-long measurement, allowing us to compensate for slow (sub-Hz) flux drift and improve resolution of the phonon number states (Fig.~\ref{fig_SI_qubit_tracking}).

We note that fluctuations in the qubit frequency cause small ($\sim 1.5\%$) changes in the qubit-phonon detuning $\Delta = \omegage - \omegam$, which in turn cause dispersion in $\chi$. Namely, for small changes $\delta\Delta$ in the detuning, we expect variations $\delta\chi$ in the dispersive shift per phonon of order
\begin{equation}
    \frac{\delta\chi}{\chi} \simeq -\left(\frac{2\Delta -\alpha}{\Delta - \alpha}\right)\frac{\delta\Delta}{\Delta}\,.
    \label{eq_chi_dispersion}
\end{equation}
Over the duration of the number splitting measurements, we estimate that the peak-to-peak splitting varies by up to $75~\text{kHz}$ ($\delta\chi/\chi \approx 2.5\%$) given our operating parameters. Although this effect is small, it is preferable to reduce flux noise by more direct measures --- post-processing can only improve the spectral clarity of phonon number peaks to the extent that $\delta\chi \ll \gamma, \kappa$. One solution is to move to fixed-frequency qubits and use the ac-Stark shift for frequency control as done in other quantum acoustics experiments~\cite{Chu2018}.

\subsection{Supplementary Data}
\begin{figure}[t!]
    \centering
    \includegraphics[width=89mm]{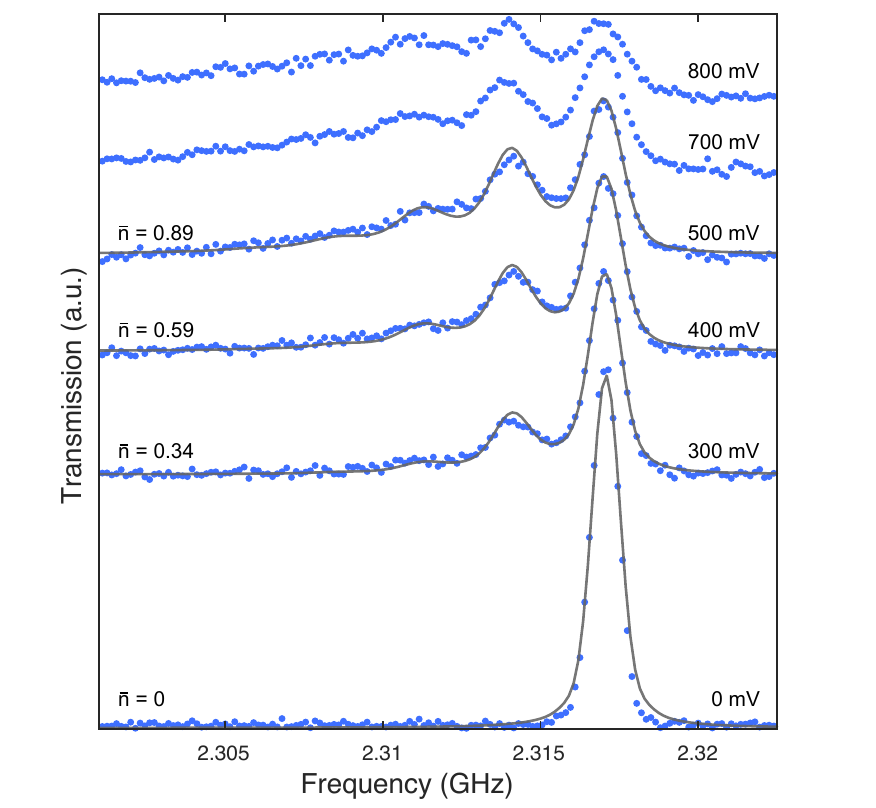}
    \caption{{\bf Phonon number splitting, supplementary dataset.}\label{fig:supplementary_data}}
\end{figure}
We made another set of phonon-number splitting measurement, the results of which are shown in Fig.~\ref{fig:supplementary_data}. These measurements were done with the same mechanical resonator. As before, the shown fits have only one fit parameter, the correction factor for the mechanical drive amplitude. For this data set, we found that for the first four voltage drive amplitudes, a correction factor of $1.0$ led to curves that fit our observation. The higher voltage data fit better to a correction factor of $0.78$ (not shown). The cause of this discrepancy is unknown, and it may have arisen from some change in the experimental setup.

\section{Numerical simulations of qubit spectra}
\label{sec_numerical_sims}
\subsection{System Hamiltonian}
We model the transmon as a nonlinear mode at frequency $\omegage$ with anharmonicity $\alpha$, and the mechanical oscillator as a linear mode at frequency $\omegam$. The two systems are coupled linearly at a rate $g$ and are driven via the transmon at a time-dependent Rabi rate $\Omega(t)$. In a frame rotating at the drive frequency $\omegad$, the total Hamiltonian can be written as $\hat{H} = \hat{H}_0 + \hat{H}_\text{d}$, with
\begin{align}
    \label{eq_full_hamiltonian}
    \hat{H}_0 &= \Deltage \opad \opa - \frac{\alpha}{2}\opad\opad\opa\opa + \Deltam \opbd\opb + g(\opa\opbd + \opad\opb), \\
    \label{eq_full_hamiltonian_drive}
    \hat{H}_\text{d} &= \frac{1}{2}(\Omega (t)\opa + \Omega^*(t)\opad),
\end{align}
where $\opa$ and $\opb$ are the annihilation operators for the transmon and mechanical modes, respectively, and $\Delta_i := \omega_i - \omegad$. While transforming to this frame, counter-rotating terms of the form $\opa\opb e^{-2i\omegad t}$ are neglected. In the dispersive limit, where both $g^2/\Delta^2 \ll 1$ and $g^2/(\Delta - \alpha)^2 \ll 1$, with $\Delta := \omegage - \omegam$, Eq.~\ref{eq_full_hamiltonian} reduces to an effective Hamiltonian
\begin{equation}
    \label{eq_dispersive_hamiltonian_SI}
    \hat{H}_{0, \text{eff}} = \Deltage \opad\opa + \Deltam \opbd\opb + 2\chi \opad \opa \opbd \opb
\end{equation}
that is diagonal in the number basis of both systems \cite{Koch2007}. The dispersive coupling rate $\chi$ is related to the original system parameters through
\begin{equation}
    \label{eq_dispersive_coupling}
    \chi = -\frac{g^2}{\Delta} \frac{\alpha}{\Delta - \alpha}.
\end{equation}
This effective Hamiltonian explicitly reveals the dependence of the qubit frequency on the number of phonons $\hat{n} = \opbd\opb$, with $\omegage(\hat{n}) = \omegage + 2\chi\hat{n}$. In particular, we expect to observe phonon number splitting of the qubit spectroscopic line if $|2\chi| \gg \max\{\gamma, \kappa\}$, when the splitting exceeds the largest of the qubit and mechanical linewidths ($\gamma$ and $\kappa$, respectively). In our system $2\chi/2\pi = -3.12\,\text{MHz}$, $\gamma/2\pi \approx 600\,\text{kHz}$, $\kappa/2\pi = 370\,\text{kHz}$, and $(g/\Delta)^2 \approx 0.03$.

\subsection{Master equation simulations}
We model our experiment by  performing time-domain master equation simulations of all the known dynamics of the system including the full pulse sequence using the QuTip package \cite{Johansson2011}. At zero temperature, the state of the system $\oprho (t)$ evolves in time according to the master equation
\begin{equation}
    \label{eq_master_equation}
    \dot{\oprho} = i[\oprho, \hat{H}] +  \frac{\gamma_\phi}{2}D[\opad\opa]\oprho + \gamma_1 D[\opa]\oprho + \kappa D[\opb]\oprho.
\end{equation}
Here $\gamma_1$ and $\gamma_\phi$ are the energy relaxation and pure dephasing rates of the qubit, respectively, $\kappa$ is the relaxation rate of the mechanical oscillator, and $D[\hat{A}]$ is the Lindblad superoperator:
\begin{equation}
    D[\hat{A}]\oprho := \hat{A}\oprho\hat{A}^\dagger - \frac{1}{2}\hat{A}^\dagger \hat{A}\oprho - \frac{1}{2}\oprho \hat{A}^\dagger \hat{A}.
\end{equation}
We initialize the system in the ground state $\oprho(0) = |g\rangle \langle g | \otimes |0\rangle \langle 0 |$  and evolve the state by numerically integrating Eq.~\ref{eq_master_equation} using the \textit{full} Hamiltonian of Eqs.~\ref{eq_full_hamiltonian},~\ref{eq_full_hamiltonian_drive}. We use the full Hamiltonian as opposed to the effective dispersive Hamiltonian (Eq.~\ref{eq_dispersive_hamiltonian_SI}) for two reasons: first, Eq.~\ref{eq_dispersive_hamiltonian_SI} is only correct in the limit $g/\Delta \ll 1$. More importantly, since we can only drive the mechanical system through the transmon, the only way to model the excitation of phonons in our device is by using the full Hamiltonian, as the dispersive Hamiltonian does not allow the two systems to exchange energy.

We first evolve the state $\oprho(t)$ over the course of the phonon excitation pulse of duration $\tau_{\text{mech}} = 175 \, \text{ns}$, setting $\omegad = \omegam$ as the drive frequency --- the exact choice of $\omegad$ is not important because the short pulse has a bandwidth on the order of several MHz. Here, the time-dependent Rabi rate of Eq.~\ref{eq_full_hamiltonian_drive} is $\Omega(t) = \Omega_0[1 - \cos(2\pi t/\tau_\text{mech})]/2$ (see Section~\ref{sec_pulse_seq} for details about the pulse sequence). We use the calibration factor $A_3$ described in Section~\ref{sec_rabi} as an initial estimate for the Rabi rate $\Omega_0(V)$ at a given drive voltage $V$. We then fine-tune $A_3$ in order to better match the measurements and simulations, effectively treating it as a free fit parameter.

At the end of the phonon pulse simulation, the mechanical oscillator is left approximately in a coherent state while the qubit is close to its ground state $\ket{g}$. Due to the small residual entanglement between them, the state of the system $\oprho(\tau_\text{mech})$ is in general non-separable. We use $\oprho(\tau_\text{mech})$ as the initial state for a subsequent simulation of the spectroscopy pulse. Here the pulse length is $\tau = 1.5 \, \mu\text{s}$, and the Rabi rate $\Omega_0(V)$ is determined using the calibration factor $A_1$. For this simulation, the drive frequency $\omegad = \omegage' + \delta$ is swept over a range of detunings $\delta/2\pi \in [-20, 10] \, \text{MHz}$ in the vicinity of the (renormalized) qubit frequency, $\omegage' = \omegage + g^2 / \Delta$. For each detuning $\delta$, we calculate the excited state population $p_\text{e}(\tau_\text{mech} + \tau) = \text{Tr} \{ \oprho_\text{f} |e\rangle \langle e| \}$, where $\oprho_\text{f} := \oprho(\tau_\text{mech} + \tau)$ is the final state at the end of the spectroscopy pulse. Noticeably, the excited state population $p_\text{e} (t)$ exhibits rapid oscillation during the qubit readout. This is caused by the strong coupling between the qubit and the mechanics in our system. However, in the actual experiment, the fast oscillations in $p_\text{e}(t)$ are filtered out due to the finite bandwidth of the readout resonator. To approximate the experimental readout, we freely evolve the system with $\hat{H}_0$ and calculate $p_\text{e} (t)$ over a time range $[\tau_\text{mech} + \tau, \tau_\text{mech} + \tau + \delta \tau]$ (with $\delta \tau = 100 \, \text{ns}$) after the end of spectroscopy pulse and record the time-averaged value $\overline{p_\text{e}(t)}$ as the actual measurement result. Time-averaged values are collected from the simulation because in the actual experiment, the readout resonator filters out fast oscillations in $p_\text{e}$. Repeating this simulation over a range of detunings produces the simulated qubit excitation spectra $p_\text{e} (\delta)$ in Fig.~4 of the main text. We remark that all parameters used in these simulations except $A_3$ were directly measured (see Table~\ref{table_device_params}) --- no free parameters were used to model the data, with the exception of an overall scaling factor and an offset.

\subsection{Excitation of phonons through the transmon}
\label{sec_phonon_excitation}

In this experiment, the mechanical system can only be driven through the transmon's charge line (see Eq.~\ref{eq_full_hamiltonian_drive}). Despite the fact that the transmon is a highly nonlinear system, it is still possible to excite many phonons at a time and prepare the mechanical oscillator in states that closely approximate coherent states. This is because in the dispersive limit $g/\Delta \ll 1$, the polaritons (the eigenvectors of the full Hamiltonian of Eq.~\ref{eq_full_hamiltonian}) are well-separated into transmon-like and phonon-like modes. More precisely, we can diagonalize the linear part of Eq.~\ref{eq_full_hamiltonian} through the polaron transformation \cite{Girvin2011}
\begin{equation}
  \left( {\begin{array}{c}
   \opc_+ \\
   \opc_- \\
  \end{array} } \right) = 
  \left( {\begin{array}{cc}
  \cos(\theta/2) & \sin(\theta/2) \\
  -\sin(\theta/2) & \cos(\theta/2)
  \end{array}} \right)
  \left( {\begin{array}{c}
  \opa \\
  \opb 
  \end{array} } \right),
\end{equation}
with $\tan(\theta) = 2g/\Delta$. In the dispersive limit, the polaritons can be written approximately as $\opc_+ \approx \opa + (g/\Delta)\opb$ and $\opc_- \approx \opb - (g/\Delta)\opa$, or alternatively the original modes can be written as $\opa \approx \opc_+ - (g/\Delta)\opc_-$ and $\opb \approx \opc_- + (g/\Delta)\opc_+$. The linear part of the diagonalized Hamiltonian (including the drive) is therefore
\begin{equation}
    \hat{H}'_\text{linear} = \Delta_+ \opcd_+ \opc_+ + \Delta_- \opcd_- \opcd_- \\ + \frac{1}{2}[\Omega(t)(\opc_+ - (g/\Delta)\opc_-) + \text{h.c.}],
\end{equation}
where $\Delta_\pm := \omega_\pm - \omegad$, and $\omega_\pm = (\omegage + \omegam)/2 \pm \sqrt{\Delta^2/4 + g^2}$ are the polariton frequencies. If we set $\omegad = \omega_-$, we are left with
\begin{equation}
    \hat{H}'_\text{linear} = \Delta_+ \opcd_+ \opcd_+ + \frac{1}{2}[\Omega(t)(\opc_+ - (g/\Delta)\opc_-) + \text{h.c.}].
\end{equation}
In this frame, the transmon-like polariton is a high-energy excitation that adiabatically follows the time-dependent drive $\Omega(t)$ if $\tau_\text{mech} \gg \Delta_+^{-1}$ --- that is, if the pulse length is much larger than the timescale associated with the transmon transition --- and the phonon-like polariton is driven directly at a suppressed rate $(g/\Delta)\Omega(t)$. In this experiment $\tau_\text{mech} = 175 \, \text{ns}$ and $\Delta_+^{-1} \approx 11 \, \text{ns}$, so the transmon rapidly responds to the drive and efficiently transfers its energy to the mechanical system.

\begin{figure*}[t!]
    \centering
    \includegraphics[width=178mm]{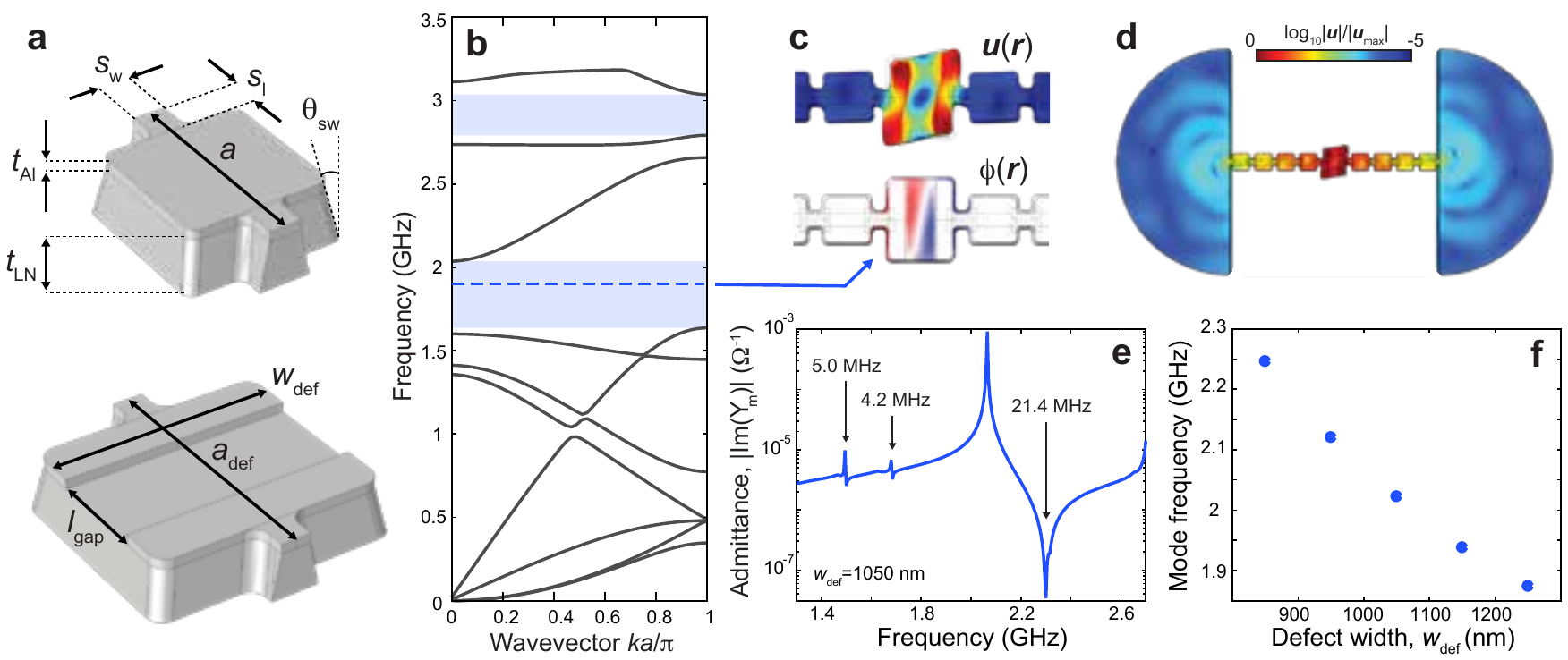}
    \caption{{\bf Phononic crystal cavity design. a,} Simulation geometries of the mirror cell (top) and defect cell (bottom). {\bf b,} Complete band diagram (including all polarizations and symmetries) of the mirror regions. A bandgap over the range $\sim [1.6, 2.0]\,\text{GHz}$ is clearly visible. For these simulations we use the same mirror cell parameters as those of the fabricated devices: $a = 1 \, \mu\text{m}$, $s_l = 330 \, \text{nm}$, $s_w = 150 \, \text{nm}$, $t_\text{Al} = 50 \, \text{nm}$, $t_\text{LN} = 230 \, \text{nm}$, and $\theta_\text{sw} = 11^\text{o}$. The discrepancy between the observed and simulated bandgap positions is not understood and could be attributed to a number of factors, including possible differences in the material constants of our LN films and those used for the simulations. {\bf c,} Deformation $\mathbf{u(r)}$ (top) and electrostatic potential $\phi(\mathbf{r})$ (bottom) of a localized defect mode at $\nu \approx 1.9 \,\text{GHz}$. Here we use $a_\text{def} = 1.3 \, \mu\text{m}$, $w_\text{def} = 1.25 \, \mu\text{m}$, and $l_\text{gap} = 500 \, \text{nm}$. The mode deformation is predominantly polarized in the plane of the phononic crystal, and the polarization generated by the piezoelectricity in LN is predominantly aligned along the direction of the electric field produced by the electrodes, as is evident by the electrostatic potential. {\bf d,} Deformation of the same mode as that in (c), with a view of the entire resonator. The color indicates $\log_{10}|\mathbf{u(r)}|/|\mathbf{u}_\text{max}|$, illustrating that even with $N = 4$ mirror cells the mode is tightly localized to the defect region. In the measured devices, $N = 8$. {\bf e,} Imaginary part of the electromechanical admittance $Y_\text{m}(\omega)$, obtained from finite-element simulations of the structure shown in {\bf d}. Using Foster synthesis we extract the coupling rates $g$ of each of the modes associated with the pole/zero pairs that are visible in the response. {\bf f,} Frequency of the strongly coupled modes, as a function of $w_\text{def}$. Their distribution (though not their absolute values) agrees fairly well with that of the observed modes.}
    \label{fig_SI_pcdc_design}
\end{figure*}

\section{Phononic crystal cavity design}
\label{sec_pcc_design}

As described in the main text, the mechanical resonators used in this work are one-dimensional phononic crystal resonators; each resonator is formed by introducing a single defect site to an artificial lattice that is patterned onto the LN. This localizes a set of vibrational modes at the defect site, provided the modes lie widthin the bandgap of the surrounding lattice. More practically, this configuration can be thought of as a wavelength-scale resonator surrounded by acoustic ``Bragg mirrors''. Each unit cell of the mirror region is comprised of a square-shaped block of LN uniformly covered by a $50 \, \text{nm}$ aluminum layer. As shown in Fig.~\ref{fig_SI_pcdc_design}a, the mirror cell is parameterized by its lattice constant $a$, strut length $s_\text{l}$, and strut width $s_\text{w}$. Additionally, there are other geometric parameters which we cannot tightly control during fabrication, such as the LN thickness $t_\text{LN}$, sidewall angle $\theta_\text{sw}$, and corner fillet radius $R$. We numerically simulate \cite{comsol2013} the eigenmodes of the mirror cell using Floquet boundary conditions, sweeping the wavevector $k$ over the first Brillouin zone $k \in [0, \pi/a]$. This produces a band diagram such as the one shown in Fig.~\ref{fig_SI_pcdc_design}b, where we used the same set of mirror cell parameters as those of the fabricated device. The diagram shows all possible bands of the structure within the frequency range of interest --- including all polarizations and symmetries --- and exhibits a clear phononic bandgap over the range $\sim[1.6, 2.0] \,\text{GHz}$. This gap is similar in size to that observed in the experiment ($\sim 400\,\text{MHz}$) but is centered at a lower frequency, which could be due to differences in the material properties of our films and those used for the simulations. As a final step, we verify the robustness of the phononic bandgap to variations in the mirror cell parameters in order to ensure that fabrication-induced fluctuations will not drastically alter the size or position of the gap.

The defect cell is created by stretching the local lattice constant to a larger value $a_\text{def} > a$ and introducing a break in the aluminum metallization, effectively forming a pair of electrodes separated by a gap $l_\text{gap}$ (Fig.~\ref{fig_SI_pcdc_design}a). This configuration supports modes that lie within the phononic bandgap and are therefore localized to the defect site (Fig. \ref{fig_SI_pcdc_design}c, d). Through the piezoelectric effect, the strain $S_{jk}$ associated with each mode induces a polarization $P_i = e_{ijk}S_{jk}$ in the crystal, where $e_{ijk}$ is the piezoelectric coupling tensor. The modes of the structure can couple strongly to the qubit if the polarization field $P$ overlaps with the electric field of the electrodes and is predominantly aligned along the same direction. Our devices are fabricated on X-cut LN, with the direction of propagation of the phononic crystals pointing along the Y crystal axis. This orientation allows for defect modes that have the `correct' polarization, as shown in Fig.~\ref{fig_SI_pcdc_design}c. Using the techniques outlined in Ref.~\cite{Arrangoiz-Arriola2016}, we calculate coupling rates $g/2\pi \approx 20-22 \, \text{MHz}$ for the fabricated defect geometries, in modest ($\delta \sim 30\,\%$) agreement with our measurements (Fig.~\ref{fig_SI_pcdc_design}e). In addition, the defects generally support other localized modes which do not couple as strongly ($g/2\pi \lesssim 5\,\text{MHz}$). 

The device used in this experiment contains an array of five resonators that have the same mirror design, but different values of the defect width $w_\text{def}$. As discussed earlier, each cavity supports a small number of localized modes (Fig.~\ref{fig_SI_pcdc_design}e), but only one of them has the correct polarization. In Fig.~\ref{fig_SI_pcdc_design}f we show the simulated frequencies of such modes for the values of $w_\text{def}$ used in our device. These simulation results clearly show that the five strongly coupled modes we observe each correspond to a separate resonator in the array, and also explain the origin of additional weakly coupled modes present in the spectrum.

\section{Optimization of the dispersive coupling $\chi$}

\begin{figure*}[t!]
    \centering
    \includegraphics[width=178mm]{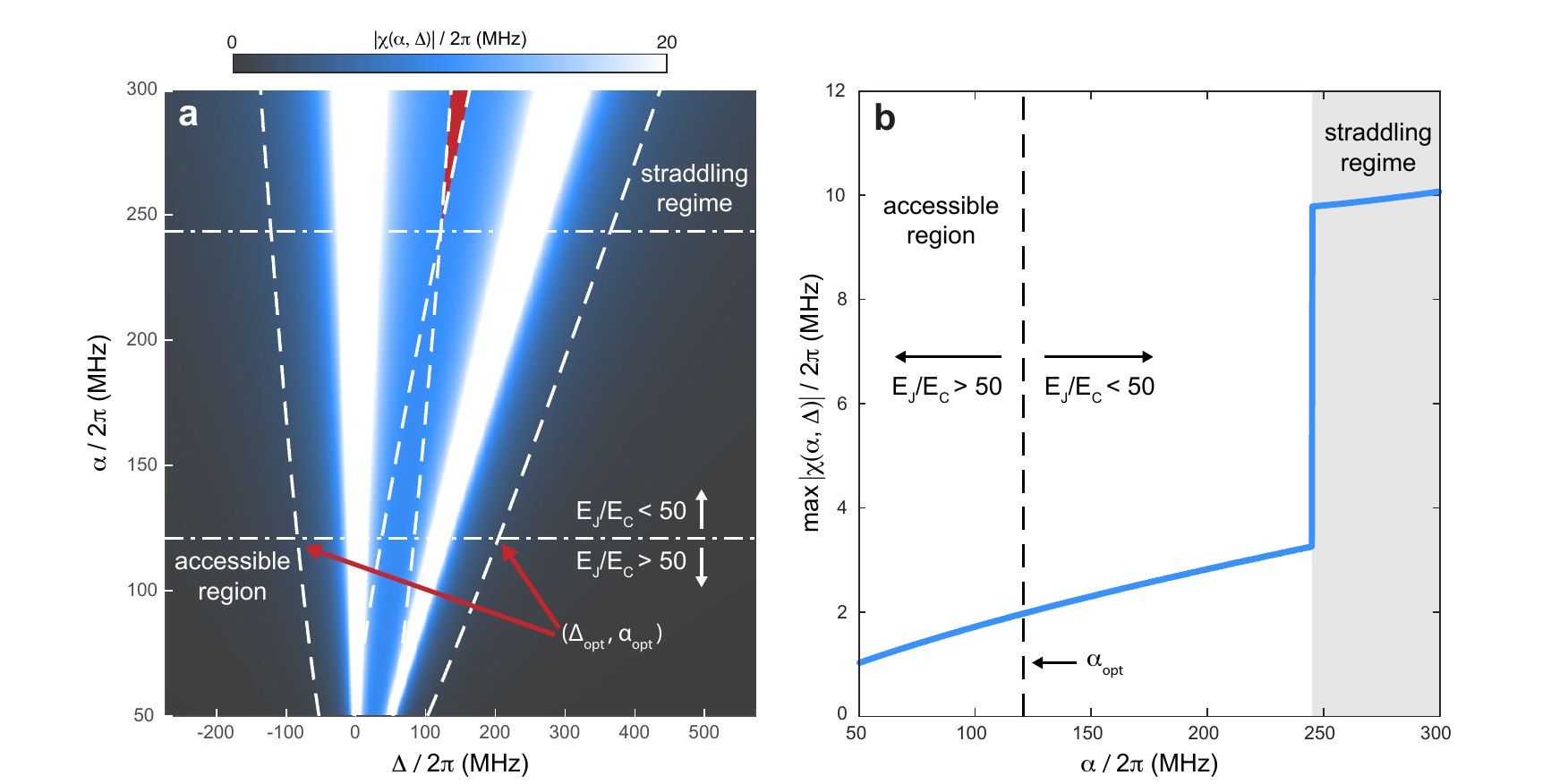}
    \caption{{\bf Dispersive coupling optimization. a,} Dispersive coupling rate $|\chi(\alpha, \Delta)|$ as a function of the detuning $\Delta = \omegage - \omegam$ and the anharmonicity $\alpha = \omegage - \omegaef$. The plotted values are bounded above at $|\chi|/2\pi = 20\, \text{MHz}$, which artificially saturates the color near the poles at $\Delta = 0$ and $|\Delta - \alpha| = 0$. The boundaries that define the forbidden regions $|\Delta| < 5g$ and $|\Delta - \alpha| < 5g$ around each pole are indicated by dashed lines. We also indicate the value of $\alpha$ above which the transmon limit condition $E_J / E_C > 50$ is violated (dash-dotted line), as well as the value of $\alpha$ above which the straddling regime $0 < \Delta < \alpha$ (shaded red) becomes accessible due to the poles becoming sufficiently separated. The optimal points $(\Delta_\text{opt}, \alpha_\text{opt})$ for our device are indicated. {\bf b,} Maximum value of $|\chi(\alpha, \Delta)|$ over all possible detunings $\Delta$, plotted as a function of $\alpha$. This value increases sublinearly for small $\alpha$ until a threshold is reached and the straddling regime becomes accessible. Above this threshold the dispersive coupling becomes much larger ($\chi/2\pi \approx 10 \, \text{MHz}$), but this region is not accessible in our device as it violates the constraint $E_J / E_C > 50$.}
    \label{fig_SI_chi_optimization}
\end{figure*}

In addition to designing phononic crystal defect resonators with large piezoelectric coupling (see Section \ref{sec_pcc_design}), we took great care to maximize the dispersive coupling $|\chi|$ between the transmon and the mehcanical oscillators by choosing an appropriate set of transmon parameters. The transmon qubit used in this work is of the `Xmon' style, with all aspects of its design (including capacitor shape, SQUID loop, charge line, flux line, and resonator coupling capacitor) closely following those outlined in Ref.~\cite{Barends2013}. Since $\chi$ is a function of the coupling rate $g$, the anharmonicity $\alpha = \omegage - \omega_{\text{ef}}$, and the detuning $\Delta = \omegage - \omegam$ (see Eq.~\ref{eq_dispersive_coupling}), it may seem that in order to maximize $|\chi|$, we must do so by varying all three parameters $(g, \alpha, \Delta)$. However, once we constrain the transmon frequency (to a value in the region where we expect to see mechanical modes, $\sim 2.0 - 2.4\,\text{GHz}$) and the defect resonator design is fixed, $g$ is no longer a free parameter --- rather it is determined entirely by $\alpha$. To see this, we write $g$ in terms of the Josephson energy $E_J$ and charging energy $E_C$ of the transmon, as well as the Foster circuit values $(C_0, C_1, L_1)$ that encode the properties of the defect resonator (see Ref.~\cite{Arrangoiz-Arriola2016} for details):
\begin{equation}
    \hbar g = 8E_C^{\phi, \theta}n_\text{zp}^{\theta}n_\text{zp}^{\phi},
\end{equation}
where 
\begin{align}
    E_C^{\phi, \theta} &= \frac{C_0}{C_1 + C_\Sigma}\frac{e^2}{2[C_0 + (C_1^{-1} + C_\Sigma^{-1})^{-1}]}, \\
    n^\phi_\text{zp} &= \frac{1}{2}\left(\frac{E_J}{2E_C}\right)^{1/4}, \\
    n^\theta_\text{zp} &\approx \frac{1}{2}\left(\frac{C_0 + C_1}{L_1} \frac{\Phi_0^2}{e^2}\right)^{1/4},
\end{align}
$C_\Sigma = e^2/2E_C$ is the transmon capacitance, and $\Phi_0 = \hbar/2e$ is the reduced flux quantum. After some algebra, we obtain $\hbar g = \xi (8E_J E_C^3)^{1/4}$, where
\begin{equation}
    \xi = \frac{C_0}{C_0 + C_1}\left(\frac{C_0 + C_1}{ L_1}\frac{\Phi_0^2}{e^2}\right)^{1/4} = \text{const.}
\end{equation}
The value of the dimensionless coupling constant $\xi$ is determined entirely by the defect resonator design and is independent of the transmon parameters. For our system $\xi \approx 0.037$ --- see Section \ref{sec_pcc_design} for details on how the Foster circuit values were calculated. The final constraint comes from the transmon energy-level spacing~\cite{Koch2007}
\begin{equation}
    \label{eq_transmon_frequency}
    \hbar\omegage = \sqrt{8 E_J E_C} - E_C.
\end{equation}
From this equation, if $\omegage$ and $E_C$ are both fixed, then $E_J$ is constrained to obey $E_J = (\hbar \omegage + E_C)^2/8E_C$. Therefore, we obtain an expression for $g$ in terms of the only remaining free design parameter, $\alpha = E_C/\hbar$:
\begin{equation}
    g(\alpha) = \xi \sqrt{\alpha (\omegage +\alpha)}.
\end{equation}
The optimization problem can then be written as 
\begin{align}
    \text{arg}\, \text{max}\,\,\,\,\, &|\chi(\alpha, \Delta)| \\
    \label{eq_dispersive_constraint_1}
    \text{s.t.} \,\,\,\,\, &|\Delta| \geq 5g(\alpha), \\
    \label{eq_dispersive_constraint_2}
    &|\Delta - \alpha| \geq 5g(\alpha), \\
    \label{eq_transmon_limit_constraint}
    &\frac{(\omegage + \alpha)^2}{8\alpha^2} \geq 50.
\end{align}
The first two constraints (Eqs. \ref{eq_dispersive_constraint_1}, \ref{eq_dispersive_constraint_2}) are imposed to ensure the system is in the dispersive limit ($g^2/\Delta^2 \ll 1$ and $g^2/(\Delta - \alpha)^2 \ll 1$ ), while the third constraint (Eq.~\ref{eq_transmon_limit_constraint}) is imposed to ensure the qubit is in the transmon limit ($E_J/E_C \gg 1$). Solving this optimization problem numerically (with $\omegage/2\pi = 2.3\,\text{GHz}$) we obtain the optimal values $\alpha_\text{opt}/2\pi \approx 120\,\text{MHz}$ and $\Delta_\text{opt} = -5g$ (or alternatively $\Delta_\text{opt} - \alpha = 5g$), with $g(\alpha_\text{opt})/2\pi \approx 22\,\text{MHz}$. It is interesting to note that the coupling constant $\xi$ is so large in our system that --- under the imposed constraints --- the straddling regime $0 < \Delta < \alpha$ is not accessible (where $|\chi|$ would be much larger). This is because within the allowed range of $\alpha$ values, $g(\alpha)$ is so large that the entire straddling region violates the dispersive constraints (Eqs.~\ref{eq_dispersive_constraint_1},~\ref{eq_dispersive_constraint_2}). Since $g(\alpha)$ grows sub-linearly, it is possible to choose a sufficiently large value of $\alpha$ such that these constraints are no longer violated, but then we violate the transmon limit constraint (Eq.~\ref{eq_transmon_limit_constraint}). See Fig.~\ref{fig_SI_chi_optimization} for a visual explanation.

\begin{thebibliography}{10}
	\expandafter\ifx\csname url\endcsname\relax
	\def\url#1{\texttt{#1}}\fi
	\expandafter\ifx\csname urlprefix\endcsname\relax\def\urlprefix{URL }\fi
	\providecommand{\bibinfo}[2]{#2}
	\providecommand{\eprint}[2][]{\url{#2}}
	
	\bibitem{Brune1994}
	\bibinfo{author}{Brune, M.} \emph{et~al.}
	\newblock \bibinfo{title}{{From Lamb shift to light shifts: Vacuum and
			subphoton cavity fields measured by atomic phase sensitive detection}}.
	\newblock \emph{\bibinfo{journal}{Physical Review Letters}}
	\textbf{\bibinfo{volume}{72}}, \bibinfo{pages}{3339--3342}
	(\bibinfo{year}{1994}).
	\newblock \urlprefix\url{https://link.aps.org/doi/10.1103/PhysRevLett.72.3339}.
	
	\bibitem{Bertet2002}
	\bibinfo{author}{Bertet, P.} \emph{et~al.}
	\newblock \bibinfo{title}{{Direct Measurement of the Wigner Function of a
			One-Photon Fock State in a Cavity}}.
	\newblock \emph{\bibinfo{journal}{Physical Review Letters}}
	\textbf{\bibinfo{volume}{89}}, \bibinfo{pages}{200402}
	(\bibinfo{year}{2002}).
	\newblock
	\urlprefix\url{https://link.aps.org/doi/10.1103/PhysRevLett.89.200402}.
	
	\bibitem{Schuster2007}
	\bibinfo{author}{Schuster, D.~I.} \emph{et~al.}
	\newblock \bibinfo{title}{{Resolving photon number states in a superconducting
			circuit}}.
	\newblock \emph{\bibinfo{journal}{Nature}} \textbf{\bibinfo{volume}{445}},
	\bibinfo{pages}{515--518} (\bibinfo{year}{2007}).
	\newblock \urlprefix\url{http://www.nature.com/articles/nature05461}.
	
	\bibitem{Braginsky1996}
	\bibinfo{author}{Braginsky, V.~B.} \& \bibinfo{author}{Khalili, F.~Y.}
	\newblock \bibinfo{title}{{Quantum nondemolition measurements: the route from
			toys to tools}}.
	\newblock \emph{\bibinfo{journal}{Reviews of Modern Physics}}
	\textbf{\bibinfo{volume}{68}}, \bibinfo{pages}{1--11} (\bibinfo{year}{1996}).
	\newblock \urlprefix\url{https://link.aps.org/doi/10.1103/RevModPhys.68.1}.
	
	\bibitem{Ofek2016}
	\bibinfo{author}{Ofek, N.} \emph{et~al.}
	\newblock \bibinfo{title}{{Extending the lifetime of a quantum bit with error
			correction in superconducting circuits}}.
	\newblock \emph{\bibinfo{journal}{Nature}} \textbf{\bibinfo{volume}{536}},
	\bibinfo{pages}{441} (\bibinfo{year}{2016}).
	\newblock \urlprefix\url{http://dx.doi.org/10.1038/nature18949}.
	\newblock \eprint{1602.04768}.
	
	\bibitem{Aspelmeyer2014}
	\bibinfo{author}{Aspelmeyer, M.}, \bibinfo{author}{Kippenberg, T.~J.} \&
	\bibinfo{author}{Marquardt, F.}
	\newblock \bibinfo{title}{{Cavity optomechanics}}.
	\newblock \emph{\bibinfo{journal}{Reviews of Modern Physics}}
	\textbf{\bibinfo{volume}{86}}, \bibinfo{pages}{1391--1452}
	(\bibinfo{year}{2014}).
	\newblock \urlprefix\url{http://link.aps.org/doi/10.1103/RevModPhys.86.1391}.
	
	\bibitem{OConnell2010}
	\bibinfo{author}{O'Connell, A.~D.} \emph{et~al.}
	\newblock \bibinfo{title}{{Quantum ground state and single-phonon control of a
			mechanical resonator}}.
	\newblock \emph{\bibinfo{journal}{Nature}} \textbf{\bibinfo{volume}{464}},
	\bibinfo{pages}{697--703} (\bibinfo{year}{2010}).
	\newblock \urlprefix\url{http://dx.doi.org/10.1038/nature08967}.
	
	\bibitem{Gustafsson2014a}
	\bibinfo{author}{Gustafsson, M.~V.} \emph{et~al.}
	\newblock \bibinfo{title}{{Propagating phonons coupled to an artificial atom}}.
	\newblock \emph{\bibinfo{journal}{Science}} \textbf{\bibinfo{volume}{346}},
	\bibinfo{pages}{207--211} (\bibinfo{year}{2014}).
	\newblock
	\urlprefix\url{http://www.sciencemag.org/cgi/doi/10.1126/science.1257219}.
	
	\bibitem{Cohen2015}
	\bibinfo{author}{Cohen, J.~D.} \emph{et~al.}
	\newblock \bibinfo{title}{{Phonon counting and intensity interferometry of a
			nanomechanical resonator}}.
	\newblock \emph{\bibinfo{journal}{Nature}} \textbf{\bibinfo{volume}{520}},
	\bibinfo{pages}{522--525} (\bibinfo{year}{2015}).
	\newblock \urlprefix\url{http://www.nature.com/articles/nature14349}.
	
	\bibitem{Riedinger2016a}
	\bibinfo{author}{Riedinger, R.} \emph{et~al.}
	\newblock \bibinfo{title}{{Non-classical correlations between single photons
			and phonons from a mechanical oscillator}}.
	\newblock \emph{\bibinfo{journal}{Nature}} \textbf{\bibinfo{volume}{530}},
	\bibinfo{pages}{313--316} (\bibinfo{year}{2016}).
	\newblock \urlprefix\url{http://www.nature.com/doifinder/10.1038/nature16536}.
	
	\bibitem{Chu2018}
	\bibinfo{author}{Chu, Y.} \emph{et~al.}
	\newblock \bibinfo{title}{{Creation and control of multi-phonon Fock states in
			a bulk acoustic-wave resonator}}.
	\newblock \emph{\bibinfo{journal}{Nature}} \textbf{\bibinfo{volume}{563}},
	\bibinfo{pages}{666--670} (\bibinfo{year}{2018}).
	\newblock \urlprefix\url{http://www.nature.com/articles/s41586-018-0717-7}.
	
	\bibitem{Satzinger2018}
	\bibinfo{author}{Satzinger, K.~J.} \emph{et~al.}
	\newblock \bibinfo{title}{{Quantum control of surface acoustic-wave phonons}}.
	\newblock \emph{\bibinfo{journal}{Nature}} \textbf{\bibinfo{volume}{563}},
	\bibinfo{pages}{661--665} (\bibinfo{year}{2018}).
	\newblock \urlprefix\url{http://arxiv.org/abs/1804.07308
		http://www.nature.com/articles/s41586-018-0719-5}.
	\newblock \eprint{1804.07308}.
	
	\bibitem{Viennot2018}
	\bibinfo{author}{Viennot, J.~J.}, \bibinfo{author}{Ma, X.} \&
	\bibinfo{author}{Lehnert, K.~W.}
	\newblock \bibinfo{title}{{Phonon-Number-Sensitive Electromechanics}}.
	\newblock \emph{\bibinfo{journal}{Physical Review Letters}}
	\textbf{\bibinfo{volume}{121}}, \bibinfo{pages}{183601}
	(\bibinfo{year}{2018}).
	\newblock
	\urlprefix\url{https://link.aps.org/doi/10.1103/PhysRevLett.121.183601}.
	
	\bibitem{Mabuchi2002}
	\bibinfo{author}{Mabuchi, H.}
	\newblock \bibinfo{title}{{Cavity Quantum Electrodynamics: Coherence in
			Context}}.
	\newblock \emph{\bibinfo{journal}{Science}} \textbf{\bibinfo{volume}{298}},
	\bibinfo{pages}{1372--1377} (\bibinfo{year}{2002}).
	\newblock
	\urlprefix\url{http://www.sciencemag.org/cgi/doi/10.1126/science.1078446}.
	
	\bibitem{Devoret2013}
	\bibinfo{author}{Devoret, M.~H.} \& \bibinfo{author}{Schoelkopf, R.~J.}
	\newblock \bibinfo{title}{{Superconducting Circuits for Quantum Information: An
			Outlook}}.
	\newblock \emph{\bibinfo{journal}{Science}} \textbf{\bibinfo{volume}{339}},
	\bibinfo{pages}{1169--1174} (\bibinfo{year}{2013}).
	\newblock \urlprefix\url{http://www.sciencemag.org/content/339/6124/1169.full}.
	
	\bibitem{Chu2017}
	\bibinfo{author}{Chu, Y.} \emph{et~al.}
	\newblock \bibinfo{title}{{Quantum acoustics with superconducting qubits}}.
	\newblock \emph{\bibinfo{journal}{Science}} \textbf{\bibinfo{volume}{358}},
	\bibinfo{pages}{199--202} (\bibinfo{year}{2017}).
	\newblock
	\urlprefix\url{http://www.sciencemag.org/lookup/doi/10.1126/science.aao1511}.
	
	\bibitem{Thompson2008}
	\bibinfo{author}{Thompson, J.~D.} \emph{et~al.}
	\newblock \bibinfo{title}{{Strong dispersive coupling of a high-finesse cavity
			to a micromechanical membrane}}.
	\newblock \emph{\bibinfo{journal}{Nature}} \textbf{\bibinfo{volume}{452}},
	\bibinfo{pages}{72--75} (\bibinfo{year}{2008}).
	\newblock \urlprefix\url{http://www.nature.com/articles/nature06715}.
	\newblock \eprint{0707.1724}.
	
	\bibitem{Miao2009}
	\bibinfo{author}{Miao, H.}, \bibinfo{author}{Danilishin, S.},
	\bibinfo{author}{Corbitt, T.} \& \bibinfo{author}{Chen, Y.}
	\newblock \bibinfo{title}{{Standard Quantum Limit for Probing Mechanical Energy
			Quantization}}.
	\newblock \emph{\bibinfo{journal}{Physical Review Letters}}
	\textbf{\bibinfo{volume}{103}}, \bibinfo{pages}{100402}
	(\bibinfo{year}{2009}).
	\newblock
	\urlprefix\url{http://link.aps.org/doi/10.1103/PhysRevLett.103.100402}.
	
	\bibitem{Ludwig2012}
	\bibinfo{author}{Ludwig, M.}, \bibinfo{author}{Safavi-Naeini, A.~H.},
	\bibinfo{author}{Painter, O.} \& \bibinfo{author}{Marquardt, F.}
	\newblock \bibinfo{title}{{Enhanced Quantum Nonlinearities in a Two-Mode
			Optomechanical System}}.
	\newblock \emph{\bibinfo{journal}{Physical Review Letters}}
	\textbf{\bibinfo{volume}{109}}, \bibinfo{pages}{063601}
	(\bibinfo{year}{2012}).
	\newblock
	\urlprefix\url{http://journals.aps.org/prl/abstract/10.1103/PhysRevLett.109.063601
		https://link.aps.org/doi/10.1103/PhysRevLett.109.063601}.
	\newblock \eprint{1202.0532}.
	
	\bibitem{Koch2007}
	\bibinfo{author}{Koch, J.} \emph{et~al.}
	\newblock \bibinfo{title}{{Charge-insensitive qubit design derived from the
			Cooper pair box}}.
	\newblock \emph{\bibinfo{journal}{Physical Review A}}
	\textbf{\bibinfo{volume}{76}}, \bibinfo{pages}{042319}
	(\bibinfo{year}{2007}).
	\newblock \urlprefix\url{http://link.aps.org/doi/10.1103/PhysRevA.76.042319}.
	
	\bibitem{Brune1990}
	\bibinfo{author}{Brune, M.}, \bibinfo{author}{Haroche, S.},
	\bibinfo{author}{Lefevre, V.}, \bibinfo{author}{Raimond, J.~M.} \&
	\bibinfo{author}{Zagury, N.}
	\newblock \bibinfo{title}{{Quantum nondemolition measurement of small photon
			numbers by Rydberg-atom phase-sensitive detection}}.
	\newblock \emph{\bibinfo{journal}{Physical Review Letters}}
	\textbf{\bibinfo{volume}{65}}, \bibinfo{pages}{976--979}
	(\bibinfo{year}{1990}).
	\newblock \urlprefix\url{https://link.aps.org/doi/10.1103/PhysRevLett.65.976}.
	
	\bibitem{Ioffe2004}
	\bibinfo{author}{Ioffe, L.~B.}, \bibinfo{author}{Geshkenbein, V.~B.},
	\bibinfo{author}{Helm, C.} \& \bibinfo{author}{Blatter, G.}
	\newblock \bibinfo{title}{{Decoherence in superconducting quantum bits by
			phonon radiation}}.
	\newblock \emph{\bibinfo{journal}{Physical Review Letters}}
	\textbf{\bibinfo{volume}{93}}, \bibinfo{pages}{1--4} (\bibinfo{year}{2004}).
	
	\bibitem{Barends2013}
	\bibinfo{author}{Barends, R.} \emph{et~al.}
	\newblock \bibinfo{title}{{Coherent josephson qubit suitable for scalable
			quantum integrated circuits}}.
	\newblock \emph{\bibinfo{journal}{Physical Review Letters}}
	\textbf{\bibinfo{volume}{111}} (\bibinfo{year}{2013}).
	
	\bibitem{Gambetta2006}
	\bibinfo{author}{Gambetta, J.} \emph{et~al.}
	\newblock \bibinfo{title}{{Qubit-photon interactions in a cavity:
			Measurement-induced dephasing and number splitting}}.
	\newblock \emph{\bibinfo{journal}{Physical Review A}}
	\textbf{\bibinfo{volume}{74}}, \bibinfo{pages}{042318}
	(\bibinfo{year}{2006}).
	\newblock \urlprefix\url{https://link.aps.org/doi/10.1103/PhysRevA.74.042318}.
	
	\bibitem{Schuster2005}
	\bibinfo{author}{Schuster, D.~I.} \emph{et~al.}
	\newblock \bibinfo{title}{{ac Stark Shift and Dephasing of a Superconducting
			Qubit Strongly Coupled to a Cavity Field}}.
	\newblock \emph{\bibinfo{journal}{Physical Review Letters}}
	\textbf{\bibinfo{volume}{94}}, \bibinfo{pages}{123602}
	(\bibinfo{year}{2005}).
	\newblock
	\urlprefix\url{https://link.aps.org/doi/10.1103/PhysRevLett.94.123602}.
	
	\bibitem{Safavi-Naeini2011}
	\bibinfo{author}{Safavi-Naeini, A.~H.} \& \bibinfo{author}{Painter, O.}
	\newblock \bibinfo{title}{{Proposal for an optomechanical traveling wave
			phonon-photon translator}}.
	\newblock \emph{\bibinfo{journal}{New Journal of Physics}}
	\textbf{\bibinfo{volume}{13}} (\bibinfo{year}{2011}).
	
	\bibitem{Bochmann2013}
	\bibinfo{author}{Bochmann, J.}, \bibinfo{author}{Vainsencher, A.},
	\bibinfo{author}{Awschalom, D.~D.} \& \bibinfo{author}{Cleland, A.~N.}
	\newblock \bibinfo{title}{{Nanomechanical coupling between microwave and
			optical photons}}.
	\newblock \emph{\bibinfo{journal}{Nature Physics}}
	\textbf{\bibinfo{volume}{9}}, \bibinfo{pages}{712--716}
	(\bibinfo{year}{2013}).
	\newblock \urlprefix\url{http://www.nature.com/doifinder/10.1038/nphys2748}.
	
	\bibitem{MacCabe2019}
	\bibinfo{author}{MacCabe, G.~S.} \emph{et~al.}
	\newblock \bibinfo{title}{{Phononic bandgap nano-acoustic cavity with ultralong
			phonon lifetime}} \bibinfo{pages}{1--43} (\bibinfo{year}{2019}).
	\newblock \urlprefix\url{http://arxiv.org/abs/1901.04129}.
	\newblock \eprint{1901.04129}.
	
	\bibitem{Pechal2018}
	\bibinfo{author}{Pechal, M.}, \bibinfo{author}{Arrangoiz-Arriola, P.} \&
	\bibinfo{author}{Safavi-Naeini, A.~H.}
	\newblock \bibinfo{title}{{Superconducting circuit quantum computing with
			nanomechanical resonators as storage}}.
	\newblock \emph{\bibinfo{journal}{Quantum Science and Technology}}
	\textbf{\bibinfo{volume}{4}}, \bibinfo{pages}{015006} (\bibinfo{year}{2018}).
	\newblock
	\urlprefix\url{http://stacks.iop.org/2058-9565/4/i=1/a=015006?key=crossref.1783c5f155119932f902f74f614c94f6}.
	
	\bibitem{Vlastakis2013}
	\bibinfo{author}{Vlastakis, B.} \emph{et~al.}
	\newblock \bibinfo{title}{{Deterministically Encoding Quantum Information Using
			100-Photon Schrodinger Cat States}}.
	\newblock \emph{\bibinfo{journal}{Science}} \textbf{\bibinfo{volume}{342}},
	\bibinfo{pages}{607--610} (\bibinfo{year}{2013}).
	\newblock
	\urlprefix\url{http://www.sciencemag.org/cgi/doi/10.1126/science.1243289}.
	
	\bibitem{Wang2014}
	\bibinfo{author}{Wang, C.} \emph{et~al.}
	\newblock \bibinfo{title}{{Integrated high quality factor lithium niobate
			microdisk resonators.}}
	\newblock \emph{\bibinfo{journal}{Optics express}}
	\textbf{\bibinfo{volume}{22}}, \bibinfo{pages}{30924--33}
	(\bibinfo{year}{2014}).
	\newblock
	\urlprefix\url{http://www.osapublishing.org/viewmedia.cfm?uri=oe-22-25-30924{\&}seq=0{\&}html=true}.
	
	\bibitem{Dolan1977}
	\bibinfo{author}{Dolan, G.~J.}
	\newblock \bibinfo{title}{{Offset masks for lift‐off photoprocessing}}.
	\newblock \emph{\bibinfo{journal}{Applied Physics Letters}}
	\textbf{\bibinfo{volume}{31}}, \bibinfo{pages}{337--339}
	(\bibinfo{year}{1977}).
	\newblock \urlprefix\url{http://aip.scitation.org/doi/10.1063/1.89690}.
	
	\bibitem{Kelly2015}
	\bibinfo{author}{Kelly, J.}
	\newblock \emph{\bibinfo{title}{{Fault-tolerant superconducting qubits}}}.
	\newblock \bibinfo{type}{Thesis}, \bibinfo{school}{University of California,
		Santa Barbara} (\bibinfo{year}{2015}).
	
	\bibitem{Dunsworth2017}
	\bibinfo{author}{Dunsworth, A.} \emph{et~al.}
	\newblock \bibinfo{title}{{Characterization and reduction of capacitive loss
			induced by sub-micron Josephson junction fabrication in superconducting
			qubits}}.
	\newblock \emph{\bibinfo{journal}{Applied Physics Letters}}
	\textbf{\bibinfo{volume}{111}}, \bibinfo{pages}{022601}
	(\bibinfo{year}{2017}).
	\newblock \urlprefix\url{http://aip.scitation.org/doi/10.1063/1.4993577}.
	
	\bibitem{Vidal-alvarez2017}
	\bibinfo{author}{Vidal-{\'{a}}lvarez, G.}, \bibinfo{author}{Kochhar, A.} \&
	\bibinfo{author}{Piazza, G.}
	\newblock \bibinfo{title}{{Delay Lines Based on a Suspended Thin Film of X- Cut
			Lithium Niobate}}.
	\newblock \emph{\bibinfo{journal}{Ius 2017}}  (\bibinfo{year}{2017}).
	
	\bibitem{Wallraff2004}
	\bibinfo{author}{Wallraff, A.} \emph{et~al.}
	\newblock \bibinfo{title}{{Strong coupling of a single photon to a
			superconducting qubit using circuit quantum electrodynamics}}.
	\newblock \emph{\bibinfo{journal}{Nature}} \textbf{\bibinfo{volume}{431}},
	\bibinfo{pages}{162--167} (\bibinfo{year}{2004}).
	\newblock \urlprefix\url{http://www.nature.com/articles/nature02851}.
	
	\bibitem{Johansson2011}
	\bibinfo{author}{Johansson, J.~R.}, \bibinfo{author}{Nation, P.~D.} \&
	\bibinfo{author}{Nori, F.}
	\newblock \bibinfo{title}{{QuTiP: An open-source Python framework for the
			dynamics of open quantum systems}}  (\bibinfo{year}{2011}).
	\newblock \urlprefix\url{http://arxiv.org/abs/1110.0573
		http://dx.doi.org/10.1016/j.cpc.2012.02.021}.
	\newblock \eprint{1110.0573}.
	
	\bibitem{Girvin2011}
	\bibinfo{author}{Girvin, S.~M.}
	\newblock \bibinfo{title}{{Circuit QED: superconducting qubits coupled to
			microwave photons}}.
	\newblock In \bibinfo{editor}{Devoret, M.}, \bibinfo{editor}{Huard, B.},
	\bibinfo{editor}{Schoelkopf, R.} \& \bibinfo{editor}{Cugliandolo, L.} (eds.)
	\emph{\bibinfo{booktitle}{Quantum Machines: Measurement and Control of
			Engineered Quantum Systems}} (\bibinfo{publisher}{Oxford University Press},
	\bibinfo{year}{2014}).
	
	\bibitem{comsol2013}
	\bibinfo{title}{{COMSOL Multiphysics 4.4}} (\bibinfo{year}{2013}).
	
	\bibitem{Arrangoiz-Arriola2016}
	\bibinfo{author}{Arrangoiz-Arriola, P.} \& \bibinfo{author}{Safavi-Naeini,
		A.~H.}
	\newblock \bibinfo{title}{{Engineering interactions between superconducting
			qubits and phononic nanostructures}}.
	\newblock \emph{\bibinfo{journal}{Physical Review A}}
	\textbf{\bibinfo{volume}{94}}, \bibinfo{pages}{063864}
	(\bibinfo{year}{2016}).
	\newblock \urlprefix\url{https://link.aps.org/doi/10.1103/PhysRevA.94.063864}.
	
\end{thebibliography}
\end{document}